\documentclass{emulateapj}
\shorttitle{Proto-hot Jupiter Paucity}
\shortauthors{Dawson et al.}

\def\kep{\emph{Kepler\ }}
\def\kepdot{\emph{Kepler.\ }}

\def\afinal{a_{\rm final}}
\def\pfinal{P_{\rm final}}      
\def\Nsup{N_{\rm sup}}   
\def\Nmod{N_{\rm mod}}   
\def\fHJRV{f_{\rm HJ, RV}}
\def\Nmodz{N_{\rm mod, 0}}   
\def\Npp{N_{P=P\rm final}}   
\def\Nppz{N_{P=P\rm final, 0}}   
\def\barNsup{\overline{N}_{\rm sup}}   
\def\barNmod{\overline{N}_{\rm mod}}   
\def\barNmodz{\overline{N}_{\rm mod, 0}}   
\def\barNpp{\overline{N}_{P=P\rm final}}   
\def\barNppz{\overline{N}_{P=P\rm final, 0}}   
\def\Npl{N_{\rm pl}}

\def\barNpl{\overline{N}_{\rm pl}}   

\def\fHJK{f_{\rm HJ, Kepler}}

\def\rhocirc{\rho_{\rm circ}}
\def\teff{T_{\rm eff}}

\def\Pmax{P_{\rm max}}
\def\tsurvey{t_{\rm survey}}
\def\emax{e_{\rm max}}

\begin{document}
\title{The photoeccentric effect and proto-hot Jupiters III: a paucity of proto-hot Jupiters on super-eccentric orbits}
\slugcomment{Submitted to ApJ on October 26th, 2012. Accepted on October 20th, 2014.}
\author{Rebekah I. Dawson\altaffilmark{1,2,3,7}}
\author{Ruth A. Murray-Clay\altaffilmark{3,4}}
\author{John Asher Johnson\altaffilmark{3,5,6}}
\altaffiltext{1}{{Department of Astronomy, University of California, Berkeley, 501 Campbell Hall \# 3411, Berkeley CA 94720-3411}}
\altaffiltext{2}{{\tt  rdawson@berkeley.edu}}
\altaffiltext{3}{Harvard-Smithsonian Center for Astrophysics, 60 Garden St, MS-51, Cambridge, MA 02138}
\altaffiltext{4}{Department of Physics, University of California, Santa Barbara, Broida Hall, Santa Barbara, CA 93106-9530}
\altaffiltext{5}{Division of Geological and Planetary Sciences, California Institute of Technology, 1200 East California Boulevard, MC 170-25, Pasadena, CA 91125, USA}
\altaffiltext{6}{NASA Exoplanet Science Institute (NExScI), CIT Mail Code 100-22, 770 South Wilson Avenue, Pasadena, CA 91125}
\altaffiltext{7}{Miller Fellow}

\begin{abstract}
Gas giant planets orbiting within 0.1 AU of their host stars, unlikely to have formed in situ, are evidence for planetary migration. It is debated whether the typical hot Jupiter smoothly migrated inward from its formation location through the proto-planetary disk or was perturbed by another body onto a highly eccentric orbit, which tidal dissipation subsequently shrank and circularized during close stellar passages. Socrates and collaborators predicted that the latter class of model should produce a population of super-eccentric proto-hot Jupiters readily observable by \kepdot We find a paucity of such planets in the \kep sample, inconsistent with the theoretical prediction with 96.9\% confidence.  Observational effects are unlikely to explain this discrepancy. We find that the fraction of hot Jupiters with orbital period $P > 3$~days produced by the star-planet Kozai mechanism does not exceed (at two-sigma) 44\%. Our results may indicate that disk migration is the dominant channel for producing hot Jupiters with $P > 3$~days. Alternatively, the typical hot Jupiter may have been perturbed to a high eccentricity by interactions with a planetary rather than stellar companion and began tidal circularization much interior to 1 AU after multiple scatterings. A final alternative is that tidal circularization occurs much more rapidly early in the tidal circularization process at high eccentricities than later in the process at low eccentricities, contrary to current tidal theories.
\end{abstract}

\section{Introduction}

Roughly 1\% of Sun-like stars host hot Jupiters, giant planets with small semi-major axes \citep{k14742012MM,k14742012HM,k14742012W}. Unlikely to have formed in situ \citep{metal2006R}, hot Jupiters are evidence for the prevalence of planetary migration, which may take place via interactions with the proto-planetary disk (e.g. \citealt{photo1980G,photo1997W,photo2005A,photo2008I,photo2011BK}), or other bodies in the system. One or more companions can create a hot Jupiter by perturbing a cold Jupiter onto an eccentric orbit, which tidal forces shrink and circularize during close passages to the star. Proposed mechanisms for this ``high eccentricity migration'' (HEM) include Kozai oscillations induced by a distant stellar binary companion (star-planet Kozai, e.g. \citealt{photo2003W,k14742007F,photo2012N}) or by another planet in the system (planet-planet Kozai, \citealt{k14742011NF,2011LN}), planet-planet scattering \citep[e.g.][]{metal1996R,photo2006F,2008C,photo2008F,beta2008J,2010M,photo2011N,metal2012BN,2012BP}, and secular chaos \citep{photo2011WL}. We consider interactions with other bodies in the system to also encompass gravitational perturbations preceded by disk migration (e.g. \citealt{k14742011G}).

One way to distinguish whether disk migration or HEM is dominant in setting the architecture of systems of giant planets is to search for other populations of giant planets, in addition to hot Jupiters, that may also result from HEM, including 1) failed hot Jupiters, which are stranded at high eccentricities but with periapses too large to undergo significant tidal circularization over the star's lifetime, 2) Jupiters on short-period, moderately-eccentric orbits, nearing the end of their HEM journey, and 3) proto-hot Jupiters on super-eccentric orbits in the process of HEM. Recently, \citet{photo2012SK} (S12 hereafter) suggested that, if HEM is the dominant channel for producing hot Jupiters, we should readily detect a number of super-eccentric Jupiters in the act of migrating inward. Moreover, they showed that the number of super-eccentric Jupiters can be estimated from the number of moderately-eccentric Jupiters that have similar angular momentum, based on their relative circularization rates. Based on the number of moderately-eccentric, short-period Jupiters found by other planet hunting programs (tabulated in the Exoplanet Orbit Database, EOD, by \citealt{photo2011W}), S12 predicted that the \kep\ Mission should discover 5-7 proto-hot Jupiters with eccentricities $e > 0.9$ and noted that these planets should in fact already be present in the \citet{k14742011BKB} candidate collection.  

The S12 prediction requires the steady production rate of hot Jupiters throughout the Galaxy to be represented in the observed sample, as well as several conventional assumptions, including conservation of the migrating Jupiter's angular momentum, tidal circularization under the constant time lag approximation, and the beginning of HEM at or beyond an orbital period of 2 years. This prediction is a useful, quantitative test for discerning the origin of hot Jupiters. Confirmation of the S12 prediction would reveal that hot Jupiters are placed on their close-in orbits by interactions with companions, not a disk, while a paucity of proto-hot Jupiters in the \kep\ sample would inform us that HEM is not the dominant channel, or that some aspect of our current understanding of HEM is incorrect.

Motivated by the S12 prediction, we have been using what we term the  ``photoeccentric effect'' to measure the eccentricities of Jupiter-sized planets from their transit light curves (\citealt{k14742012DJ}, DJ12 hereafter). We have validated our approach by comparing our eccentricity measurement obtained from the light curve to (when available) radial-velocity measurements, including for HD-17156-b (DJ12), Kepler-419-b (formerly KOI-1474.01, \citealt{daw14}), and KOI-889-b\footnote{In unpublished work, we measured a value $e=0.5$. \citet{2013HA} measured a value of 0.589 from SOPHIE radial-velocity measurements.}. Prior to the radial velocity follow-up, Kepler-419-b was identified by \citet{photo2012DJM} (D12 hereafter) as a transiting planet candidate with a long orbital period (69.7~days), a large eccentricity ($e = 0.81 \pm 0.10$), and transit timing variations caused by a massive outer companion. Originally uncertainty in the candidate's eccentricity made it ambiguous whether Kepler-419-b is one of the proto-hot Jupiters predicted by S12 or, alternatively, a failed-hot Jupiter beyond the reach of tidal circularization over its host star's lifetime. It was later shown to be failed-hot Jupiter \citep{daw14}.

Here we examine the entire sample of \kep Jupiters to assess whether the planets expected from HEM are present. We find with 96.9\% confidence that the putative highly-eccentric progenitors of hot Jupiters are partly or entirely missing from the \kep sample. In Section \ref{sec:expect}, we update the S12 prediction, accounting for Poisson counting uncertainties and incompleteness, and translate it into a prediction for transit light curve observables. In Section \ref{sec:obs}, we compare the prediction of Section \ref{sec:expect} to the light curve properties of candidates in the \kep\ sample and conclude that there is a paucity of proto-hot Jupiters. In Section \ref{sec:explain}, we place an upper-limit on the fraction of hot Jupiters created by stellar binaries, consider the contribution of disk migration to the hot Jupiter population, and present Monte Carlo predictions for other dynamical scenarios, finding that the paucity of proto-hot Jupiters can be compatible with HEM. We conclude (Section \ref{sec:conclude}) by outlining the theoretical and observational pathways necessary to distinguish the dominant channel for hot Jupiter creation.

\section{Updated Prediction for Number of Super-eccentric Proto-hot Jupiters and Transit Light Curve Observables}

\label{sec:expect}
In Section \ref{subsec:expect}, we derive the expected number of identifiable \kep super-eccentric proto-hot Jupiters, following S12 but using updated survey samples. We refine the S12 prediction by quantifying its uncertainty and incorporating incompleteness.  In Section \ref{subsec:predict}, we describe how to confirm or rule out the existence of super-eccentric proto-hot Jupiters using \kep\ photometry alone by recasting the prediction in terms of light curve observables.

\subsection{Expected Number of Proto-hot Jupiters with $e > 0.9$ in the \kep Sample}
\label{subsec:expect}

S12 predicted that the \kep mission should discover a number of super-eccentric, hot Jupiter progenitors in the process of high eccentricity migration (HEM).  Previously (DJ12), we showed that super-eccentric planets should be easily identifiable from their transit light curves and thus precise radial-velocity (RV) follow-up is unnecessary. This is fortunate because most \kep stars are too faint to be amenable to precise RV observations. To predict the number of super-eccentric Jupiters, S12 considered a population of proto-hot Jupiters undergoing tidal circularization along a ``track'' of constant angular momentum defined by final semi-major axis, $\afinal = a (1-e^2)$ and, related by Kepler's law, final orbital period $\pfinal = P (1-e^2)^{3/2}$ for which $a$, $e$ and $P$ are the observed semi-major axis, eccentricity, and orbital period respectively.  The number of super-eccentric Jupiters ($e>0.9, \barNsup)$ along a track of constant angular momentum is related to the number of moderately-eccentric Jupiters $(0.2 < e <0.6, \barNmod)$ along the same track by:
\begin{equation}
\label{eqn:predn}
\barNsup = \barNmod r(\emax)
\end{equation}
\noindent where the variable $\emax = \left[1-\left(\pfinal/\Pmax \right)^{2/3}\right]^{1/2}$ is set by maximum observable orbital period $\Pmax$ and  $r(\emax)$ is the ratio of time spent at super-eccentricities $(0.9 < e < \emax)$ to moderate eccentricities $(0.2 < e < 0.6)$. We place bars over $\Nsup$ and $\Nmod$ to indicate that these are mean numbers. The observationally counted numbers are sampled from Poisson distributions defined by these means. 

\begin{figure*}
\includegraphics[width=.8\textwidth]{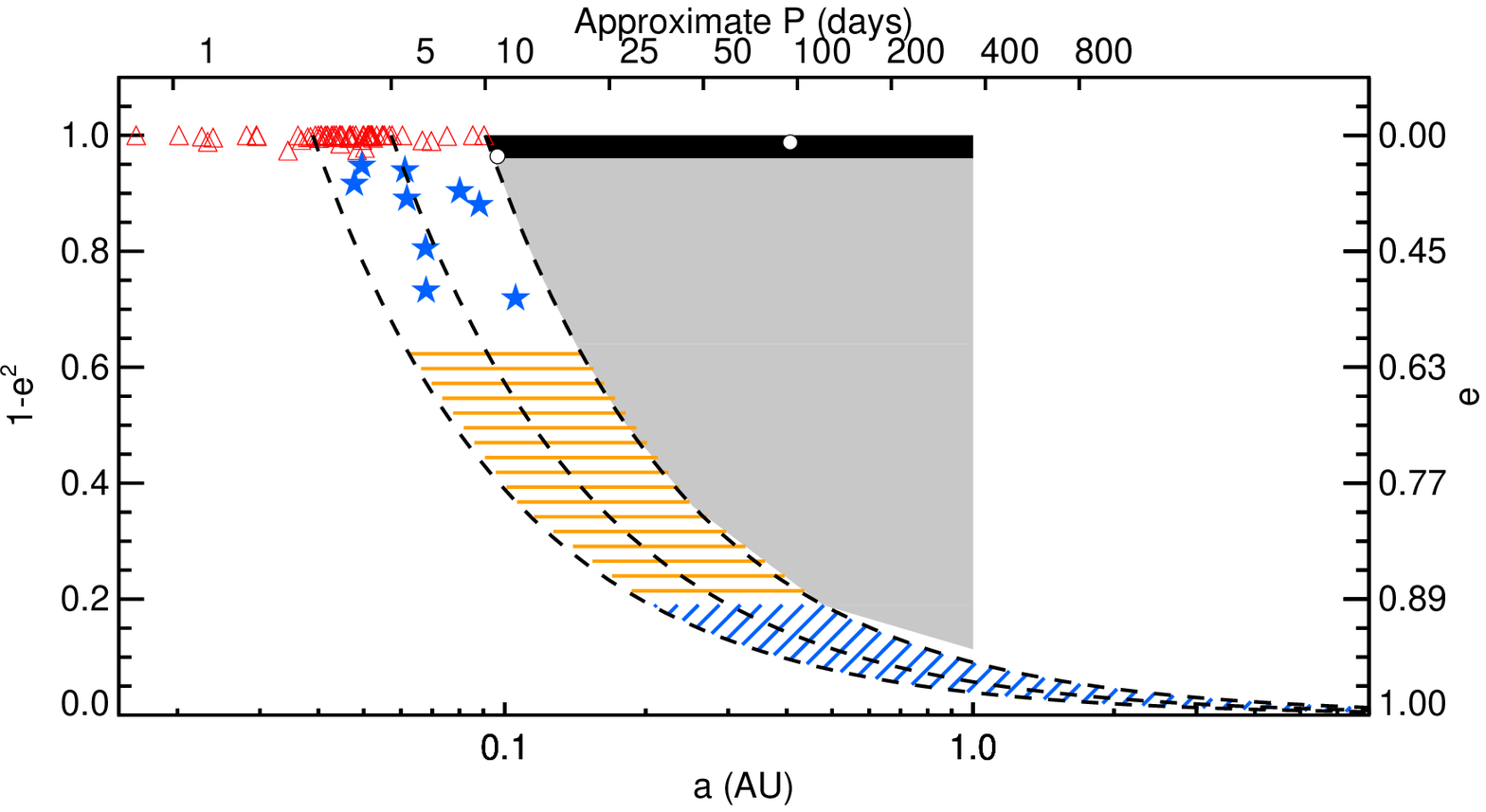}
\includegraphics[width=.8\textwidth]{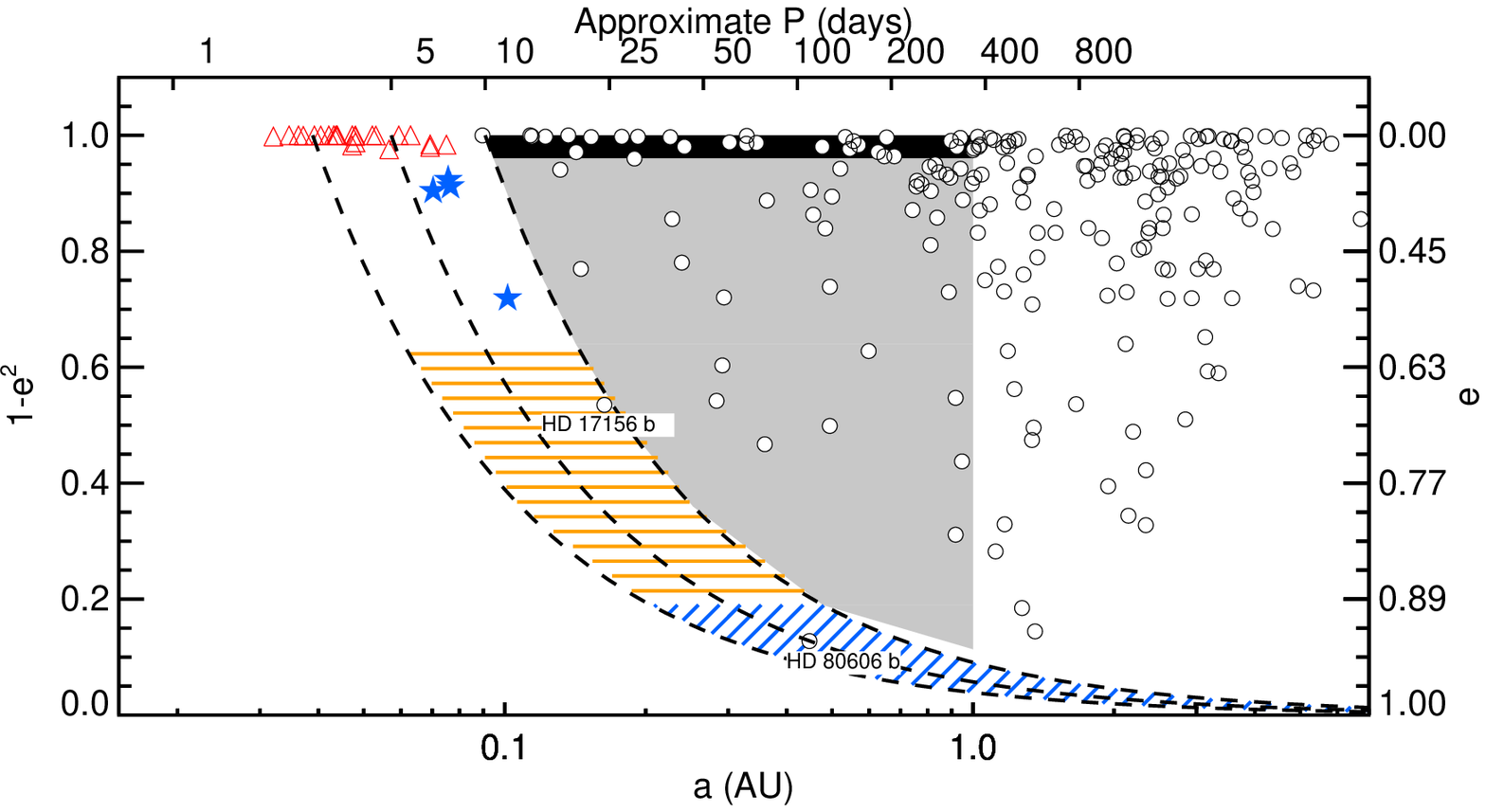}
\caption{Giant planets detected by non-\kep surveys from the EOD (\citealt{k14742012W}) using the transit technique (top) or RV technique (bottom). The top x-axis, for reference, indicates the equivalent orbital period for a planet orbiting a solar-mass star. All are Jupiter-mass $(M_p \sin i > 0.25 M_{\rm Jup}$ or $ 8 R_\oplus < R_p < 22 R_\oplus)$ planets orbiting stars with $4500 < \teff < 6500$~K, and $\log g$, within the uncertainties, consistent with $>4$. The dashed lines represent tracks of $\pfinal = 2.8,5,10$~days. The shaded and patterned regions correspond to Figure~\ref{fig:diagram}. Within the $3<\pfinal<10$~days angular momentum tracks are hot Jupiters (red triangles), moderately-eccentric Jupiters with $0.2 <e < 0.6$ (blue stars), Jupiters with $0.6 < e < 0.9$ (orange horizontal striped region), and super-eccentric Jupiters (blue, diagonal-striped region). The RV-discovered planet HD~17156~b lies in the orange, horizontal striped region, and the RV-discovered planet HD~80606~b lies in the blue, diagonal-striped region. Period valley: grey region denotes Jupiters with $\pfinal > 10$~days but interior to the uptick in giant planet frequency at 1 AU, and the black region contains circular Jupiters in this region. \label{fig:obs}}
\end{figure*}

Most Jupiters in the \kep sample lack measured eccentricities, and therefore $\barNmod$ of the \kep sample is unknown. Following S12, we make use of the sample of planets detected by non-\kep surveys, which we denote with subscript 0 (Figure~\ref{fig:obs}). To estimate $\barNmod$ along a track in the \kep sample, we use the ratio of $\barNmodz$ to the number in another class of calibration object. This other class needs to be countable in the \kep sample. Ideally, this class would be along a $\pfinal$ track. However, because the eccentricities of the \kep planets are unknown, instead the class we use is planets with orbital period $P = \pfinal$, of which there are $\barNppz$ in the calibration sample. If we assume this ratio $\barNmodz/\barNppz$ is the same for calibration sample as for the \kep sample, then we can compute the expected $\barNmod$ for the \kep sample:
\begin{equation}
\label{eqn:nmod}
\barNmod = \frac{\barNmodz}{\barNppz} \barNpp
\end{equation}

The prediction by S12 was for an ideal Kepler sample complete out to orbital periods of 2 years (Subo Dong, private communication, 2012). The \kep Mission nominally was designed to operate for three years, and operated for four years before the reaction wheels failed. To derive the expected number of super-eccentric proto-hot Jupiters in a sample of a limited timespan $\tsurvey$ we must account for incompleteness. If $N_{\rm trans, min}$ transits are the minimum number of transits required for the \kep transit pipeline to detect the proto-hot Jupiter, the completeness (with respect to this effect alone) $C_{\rm comp}$ ranges from 100$\%$ at orbital periods $\le \tsurvey/N_{\rm trans, min}$ to 0$\%$ at orbital periods of $\tsurvey/(N_{\rm trans, min}-1)$. We define $r(\emax)$ to account for incompleteness (note that the completeness is a function of eccentricity because of the one-to-one relation between eccentricity and orbital period along a given angular momentum track):
\begin{equation}
\label{eqn:rC}
r(\emax) =  \frac{\int_{0.9}^{\emax} C_{\rm comp}(e) |\dot{e}|^{-1} de}{\int_{0.2}^{ 0.6 } |\dot{e}|^{-1} de}
\end{equation}
\noindent where $\dot{e}$ due to tides raised on the planet under the constant tidal time lag approximation (\citealt{k14741998E}, \citealt{photo2010H}, S12, \citealt{2012SK,2012SKD}) and the completeness $C_{\rm comp}$(e) is
\begin{eqnarray}
\label{eqn:complete}
\left\{\begin{array}{rl}
C_{\rm comp}(e)= \nonumber\\
1,\mbox{ $e <  e_{\rm complete}$}\\
1+(1-e^2)^{3/2} (\tsurvey/\pfinal) - N_{\rm trans, min}, \\
\mbox{$e_{\rm complete}  < e < e_{\rm max}$}
\end{array} \right.\nonumber\\
\end{eqnarray}
\noindent and $e_{\rm complete} = \left[1-\left(N_{\rm trans, min}\pfinal/\tsurvey \right)^{2/3}\right]^{1/2}$,\\ \noindent and $e_{\rm max} =\left(1-\left[(N_{\rm trans, min}-1)\pfinal/\tsurvey \right]^{2/3}\right)^{1/2}$. 

Although calculations are often made under the assumption that the \kep candidate list \citep{k14742011BKB,k14742013B,metal2013BB} is complete for Jupiter-sized planets exhibiting two transits in the timespan under consideration (e.g. \citealt{metal2013F}), in this paper we estimate the completeness more conservatively.  The \kep pipeline is set up to detect objects that transit three times during the quarters over which the pipeline was run; all reported candidates that transit only 1-2 times were detected by eye (Christopher Burke and Jason Rowe, private communication, 2013). There is no estimate available for the completeness of detections made by eye. We compile a sample of candidates that is complete for giant planets that transit three times in Q1-Q16, cross-checking among several sources. We describe our sample of long-period Kepler giant planets in Appendix \ref{app:selectlong}. We employ Equation \ref{eqn:complete} using $\tsurvey = 4$ years and $N_{\rm trans, min} = 3$. We obtain $r =1.05, 0.809, 0.539$ for $\pfinal = 3, 5, 10$~days respectively. Later in this section we will update the completeness further to account for noise and missing data.

Next we describe the selection cuts we make to count $\Nmodz$ (blue stars, Figure~\ref{fig:obs}), $\Nppz$ (open symbols, Figure~\ref{fig:obs}), and $\Npp$. Because the stellar parameters from the \kep Input Catalog (KIC) are not reliable for stars outside the temperature range $4500 < \teff < 6500$~K \citep{k14742011BL} we only include stars within this temperature range in both the \kep and calibration samples\footnote{However, RV surveys are only complete out to 6100 K so the RV sample spans a more limited stellar temperature range.}. We impose a cut of stellar surface gravity $\log g > 4$ to exclude giant stars, because their KIC parameters are unreliable (we include stars with $\log g < 4$ but that are consistent with $\log g = 4$ within two sigma). All stellar parameters are taken from \citet{2013HS}.

We select planets with $8 R_\oplus < R_p < 22 R_\oplus$, where the radius is calculated using $R_p/R_\star$ from the Q1-Q12 candidate list and $R_\star$ from the updated stellar parameters in \citet{2013HS}. We follow S12 and consider two $\pfinal$ intervals: $2.8 < \pfinal < 5$ (Interval 1)\footnote{We use a lower limit of 2.8 days because 2.8 days is the $\pfinal$ below which we do not see any moderately eccentric Jupiters in the non-\kep surveys.} and $5 < \pfinal < 10$ (Interval 2). The transit probability does not change very much throughout each interval. We tabulate the counted numbers and their sources in Table~\ref{tab:count}.

Each number of counted planets (Table~\ref{tab:count}) is drawn from a Poisson distribution with an unknown mean. We wish to compute the expected number of super-eccentric proto-hot Jupiters using \emph{not} the counted numbers (which are only samples from a Poisson distribution) but rather using estimated posteriors for the mean numbers, incorporating uncertainty. We use a Jeffrey's prior. See Appendix \ref{app:poisson} for a description of our approach. Note that in the calibration sample, we exclude planets whose eccentricities are poorly constrained. For planets with $e=0$ in the EOD, we refer to the literature or fit the data ourselves and only include planets listed with $e=0$ that are constrained to have $e < 0.2$. See Appendix \ref{app:sampleshort} for more details.

\begin{deluxetable*}{llrrrl}
\tabletypesize{\footnotesize}
\tablecaption{Counted planets \label{tab:count}}
\tablewidth{0pt}
\tablehead{
 \colhead{$e$}&  \colhead{Interval } &\colhead{Counted}  &  \colhead{Mean \tablenotemark{a}}&\colhead{Sample\tablenotemark{b}}\\
 &  \colhead{[days] } \\
}
\startdata
$0.2 < e < 0.6 $		&1: 2.8-5& $\Nmodz$= 6&$\barNmodz$$=6^{+3}_{-2}  $&Cal\\
				&2: 5-10&$\Nmodz$= 7&$\barNmodz$$= 7^{+3}_{-2}  $&Cal\\
unspecified		&1: 2.8-5& $\Nppz$= 69&$\barNppz$$=69^{+9}_{-8}  $&Cal\\
				&1: 2.8-5& $\Npp$ = 24&$\barNpp$$=24\pm5$ &Kep \\
				&2: 5-10& $\Nppz$= 18 &$\barNppz$$=18^{+5}_{-4}$&Cal\\
				&2: 5-10& $\Npp $= 16 &$\barNpp$$=16\pm4$&Kep
\enddata
\tablenotetext{a}{Median, with 68.3\% confidence interval, of posterior of Poisson means, each defining a Poisson distribution from which the counted number may be sampled.}
\tablenotetext{b}{Kep = \kep; Cal = calibration non-\kep (Figure~\ref{fig:obs}).}
\end{deluxetable*}

There are two additional effects on the completeness that we now consider. First, transits may fall during gaps in the data or missing quarters. To incorporate this effect, we numerically integrate Equation \ref{eqn:rC}, inserting an extra factor $C_{\rm comp, sampled}$ into the integrand, where $C_{\rm comp, sampled}$ is the fraction of phases for which we would observe three or more transits during Q1-Q12. 

We estimate $C_{\rm comp, sampled}$ using the observation times through Q16 for the 40 hot Jupiters hosts with $2.8 < P < 10$ days in our sample (Table \ref{tab:count}). In using the hot Jupiter hosts, we assume that their observational cadence is representative of that of proto-hot Jupiter hosts. The factor $C_{\rm comp, sampled}$ automatically incorporates $C_{\rm comp}$ (Equation \ref{eqn:complete}). We require that the planet transit three times. Accounting for missing data reduces $r$ to $0.913, 0.668, 0.422$ for $\pfinal = 2.8, 5, 10$~days respectively.

Second, we consider whether the transits have sufficient signal-to-noise to be detected. A signal-to-noise ratio (SNR) of seven is the formal threshold for detection \citep{2012T,2013T}, but estimates of the current sensitivity of the \kep pipeline vary; for example, \citet{metal2013F} model the detection threshold as a ramp ranging from 0\% at SNR of eight to 100\% at SNR of sixteen. For the expected progenitors of a given hot Jupiter, the signal-to-noise is (based on \citealt{k14742012HM}, Equation 1):
\begin{equation}
{\rm S/N} = \frac{\delta}{\sigma_{\rm CDPP}} \sqrt{N_{\rm transit}\frac{t_{\rm dur}}{t_{\rm CDPP}}} 
\end{equation}
\noindent for which $N_{\rm transit} = \frac{t_{\rm survey}(1-e^2)^{3/2}}{P_{\rm HJ}}$ is the average number of transits for a hot-Jupiter progenitor with $P = P_{\rm HJ} (1-e^2)^{-3/2}$, $t_{\rm dur} =  t_{\rm HJ} /(1+e\cos\omega)$  is the duration of the progenitor's transit, $\delta$ is the transit depth, $\sigma_{\rm CDPP}$ is the combined differential photometric precision (CDPP), and $t_{\rm CDPP}$ is the timescale of the CDPP. Therefore
\begin{equation}
{\rm S/N} = \frac{\delta}{\sigma_{\rm CDPP}} \sqrt{\frac{t_{\rm survey} t_{\rm HJ} (1-e^2)^{3/2}}{P_{\rm HJ}t_{\rm CDPP}(1+e\cos\omega)}} \nonumber \\
\end{equation}
For each hot-Jupiter in our sample, we compute the most pessimistic SNR for a supereccentric progenitor: a progenitor transiting at its periapse with $e = \emax$ three times over the duration of the survey. In each case, the SNR exceeds 22, well above the 100\% detection threshold of 16 modeled by \citet{metal2013F}. Note that this approach automatically accounts for the effect of impact parameter on the transit duration by using the observed transit duration of the hot Jupiters. To get a sense for the \emph{typical} SNR of a hot-Jupiter progenitor, we compute the signal-to-noise of set of randomly generated progenitors, weighted by $\dot{e}$, the completeness (Equation \ref{eqn:complete}), and the transit probability. The resulting distribution of SNR peaks at 100, with a 15 percentile value of 54 and 1 percentile value of 28. Therefore, we expect such progenitors to be readily detectable.

\begin{figure}
\begin{centering}
\includegraphics{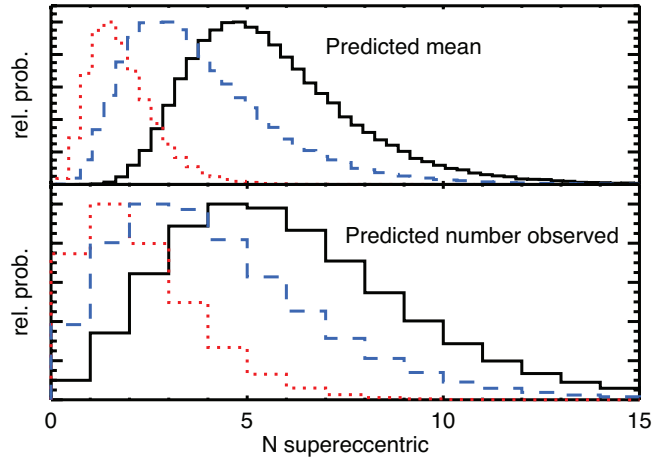} 
\caption{Top: Relative probability of predicted mean number of super-eccentric Jupiters (Interval 1: red dotted, Interval 2: blue dashed, total: black solid). Bottom: Sampling from above distribution of Poisson means to create a relative probability distribution of expected number observed.  \label{fig:exp}}
\end{centering}
\end{figure}

Next we derive the mean number of super-eccentric planets, $\barNsup$. To do so we insert the posteriors from Table~\ref{tab:count} into Equation~(\ref{eqn:predn}), making use of Equations (\ref{eqn:nmod}) and (\ref{eqn:rC}), with the $r$ listed above ($r  = 0.913, 0.668, 0.422$). We perform this procedure separately for Interval 1 and Interval 2 and use $\tsurvey = 4$ year, obtaining a $\barNsup$ posterior for each interval, which we sum to compute a total $\barNsup$ (Figure~\ref{fig:exp}). The total expected number is $\barNsup = 5.3^{+2.4}_{-1.7}$. This posterior represents a distribution of Poisson means. 

We transform the distribution of means into a distribution of expected values by sampling $\Nsup$ from $\barNsup$ according to Equation~(\ref{eqn:ppoisson}). Each sample requires first drawing a mean ($\barNsup$) from the distribution of means (Figure~\ref{fig:exp}, top row) and then drawing an observed number $\Nsup$ from the Poisson distribution with that mean. The peak of the distribution is at 4, with a 1.3\% chance of observing none.

We emphasize that after accounting for incompleteness, signal-to-noise, Poisson counting uncertainties, and updates to the observed samples, our final estimate is similar to the original S12 prediction of 5-6 supereccentric Jupiters assuming a completeness out to two year orbital periods. However, given that we will find a lack of super-eccentric Jupiters, the considerations detailed in this section will prove essential to concluding such planets are truly absent in nature rather than overlooked or coincidentally missing from this sample.

\subsection{Prediction for Transit Light Curve Observables}
\label{subsec:predict}
We expect to be able to identify super-eccentric proto-hot Jupiters in the \kep sample by fitting their transit light curves and identifying those for which the light curve model parameters are inconsistent with a circular orbit. A planet's orbital eccentricity affects its transit light curve in a number of ways (e.g. \citealt{photo2007B,2008FQ,photo2008K}). For long-period, highly eccentric, Jupiter-sized planets, the most detectable effect is on the transit duration. For a wide range of periapse orientations relative to our line of sight, a planet on a highly eccentric orbit transits its star moving at a much larger speed than if it were on a circular orbit with the same orbital period. For Jupiter-sized planets, one can distinguish the effects of the transit speed on the ingress, egress, and full transit duration from the effects of the transit impact parameter and/or limb-darkening, even with long-cadence \kep data (DJ12). Our fitting procedure fully incorporates all the uncertainty introduced by long exposure times.

For each planet, we fit a \citet{photo2002M} transit light curve model (binned to the exposure time, i.e. \citealt{photo2010Kb}) with the following parameters: the planetary-to-stellar radius ratio $R_p/R_\star$, the orbital period $P$, the inclination $i$, the scaled semi-major axis $a/R_\star$, and the quadratic limb darkening parameters $\mu_1$ and $\mu_2$. Assuming a circular orbit and a planetary mass much less than the stellar mass ($M_p << M_\star$), one can recast the scaled semi-major axis $a/R_\star$ (by substituting $a = \left[GM_\star \left(P/[2\pi]\right)^2\right]^{1/3})$ as the bulk stellar density, $\rho_\star = M_\star/(\frac{4}{3} \pi R_\star^3)$, which we will refer to simply as the stellar density hereafter. The stellar density measured from the light curve under the assumption of a circular orbit, $\rhocirc$, is related to the true stellar density, $\rho_\star$ by:
\begin{eqnarray}
\label{eqn:rhoecc}
\rho_{\star}(e,\omega)  g^{3}(e,\omega)=\rhocirc,\\
 {\rm where} \nonumber\\
g(e,\omega)=\frac{1+e\sin\omega}{\sqrt{1-e^2}}
\end{eqnarray}
is approximately the ratio of the observed transit speed (technically the transverse line-of-sight velocity) to the transit speed that the planet would have if it were on a circular orbit with the same orbital period (see \citealt{photo2010K} and DJ12 for a detailed derivation). The argument of periapse $\omega$ represents the angle on the sky plane ($\omega = 90^\circ$ for a planet transiting at periapse).

We determine $\rhocirc$ by fixing $e=0$, allowing the stellar density to vary as a free parameter in the light curve model. The resulting $\rhocirc$ is determined entirely by the shape and timing of the light curve. We then compare $\rhocirc$ to the value of $\rho_\star$ determine through other methods (i.e. stellar models fit to the temperature and surface gravity determined through colors or spectroscopy). Although $g$ is degenerate with the host star's density (Equation \ref{eqn:rhoecc}), a loose (order-of-magnitude) constraint on $\rho_\star$ is sufficient for a tight constraint on the eccentricity (DJ12), the measurement of which we will describe and perform in Section \ref{sec:obs}. In the current section, we work with the variable $\rhocirc/\rho_\star$. If $\rhocirc/\rho_\star$ is very large, then $g$ must be large, and therefore the planet is moving more quickly during transits than a planet with the same orbital period $P$ on a circular orbit. In Appendix \ref{app:star}, we summarize how our approach avoids problems caused by incorrect stellar parameters and approximations. 

\subsubsection{Expectations for Super-eccentric Planets}
\label{subsec:super}

\begin{figure*}
\includegraphics[width=\textwidth]{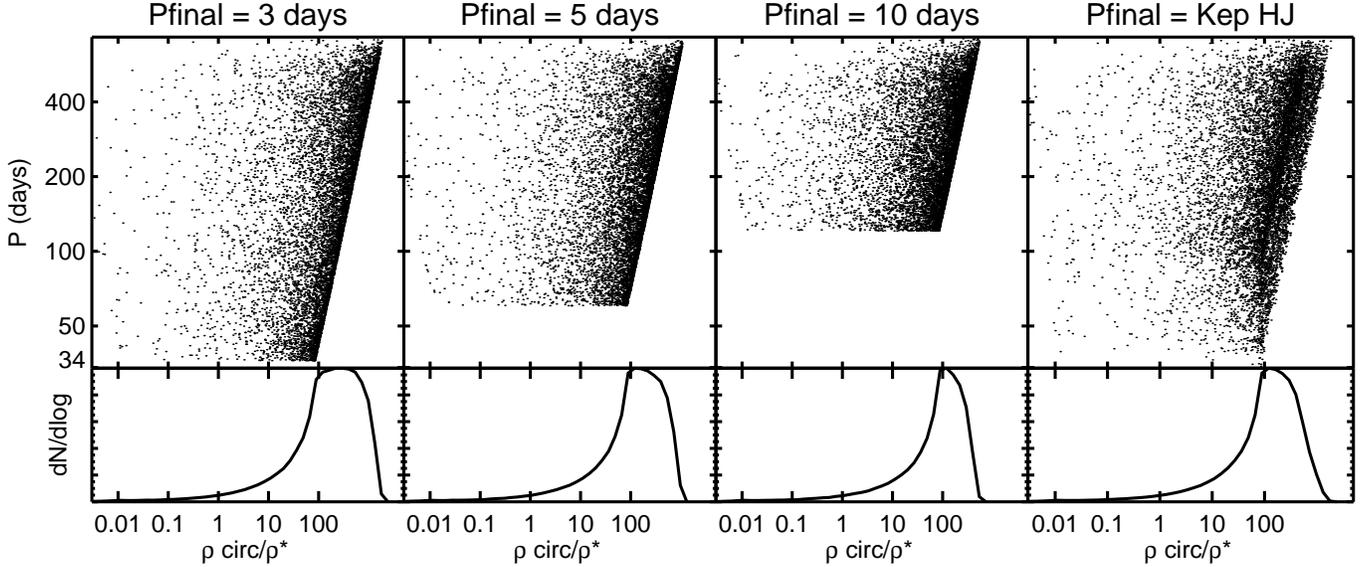}
\caption{Top: 2D posterior, orbital period $P$ vs. $\rhocirc/\rho_\star$, for planets with $e > 0.9$ and $\pfinal = 3, 5, 10$~days (panels 1-3) or $\pfinal$ drawn from \kep hot Jupiters with $2.8 < P < 10$~days (panel 4). Bottom: Posterior $\rhocirc/\rho_\star$ marginalized over orbital period. Proto-hot Jupiters with $e > 0.9$ should have anomalously large $\rhocirc$ measured from the transit light curve compared to their estimated $\rho_\star$, making them easy to identify. We expect half a dozen super-eccentric proto-hot Jupiters in the high probability density region. \label{fig:twod}} 
\end{figure*}

We perform a Monte Carlo simulation to predict the signature in the transit light curve observable $\rhocirc/\rho_\star$ expected from the super-eccentric proto-hot Jupiters (Section \ref{subsec:expect}). We generate two-dimensional (2D) probability distributions in $(P, \rhocirc/\rho_\star)$ in Figure~\ref{fig:twod}, where $P$ is the orbital period, as follows:
\begin{enumerate}
\item We begin with an assumed $\pfinal$.
\item Using the completeness Equation \ref{eqn:complete}, we generate a distribution of eccentricities $\{e_i\}$ with a normalization constant $C_{\rm norm}$ following:
\begin{equation}
{\rm Prob(e)} = \left\{\begin{array}{rl}
0&\mbox{ $e > \emax$ or $e<0.9$}\\
C_{\rm norm} C_{\rm comp, sampled} |\dot{e}|^{-1}&\mbox{$0.9< e < \emax $}
\end{array} \right.
\end{equation}
\item For each eccentricity, we compute the corresponding orbital period $P_i$ and randomly select an argument of periapse $\omega_i$. Assuming a Sun-like star, we compute the scaled semi-major axis $a_i/R_\star$.
\item We compute the transit probability:
\begin{equation}
{\rm prob}_{\rm transit} = \frac{R_\star}{a_i} \frac{1+e_i\sin \omega_i}{1-e_i^2}
\end{equation}
Then we select a uniform random number between 0 and 1. If the number is less than the transit probability, we retain $(e_i, \omega_i)$ in the distribution.
\item Then we compute $\rhocirc/\rho_\star$ using Equation~(\ref{eqn:rhoecc}).
\end{enumerate}

We use the above procedure to generate four plots, corresponding to different $\pfinal$ (Figure~\ref{fig:twod}). In the fourth panel, instead of using a single $\pfinal$, we draw the $\pfinal$ of each trial from the observed $\Npp$, weighting each $\pfinal$ by $\Nmodz/\Nppz$ in the two intervals. 

We see that a population of super-eccentric Jupiters will manifest itself as a collection of light curves with astrophysically implausible $\rhocirc$ of 10-1000 times the estimated values for $\rho_\star$. The super-eccentric proto-hot Jupiters will have orbital periods that range from $ P =  2.8 {\rm~days}/(1-0.9^2)^{3/2}=$ 34~days to two years. About 90$\%$ of the expected planets have $\rhocirc/\rho_\star > 10$, making them easy to identify. 

\subsubsection{Proto-hot Jupiters with $0.6 < e < 0.9$}

\begin{figure*}
\includegraphics[width=\textwidth]{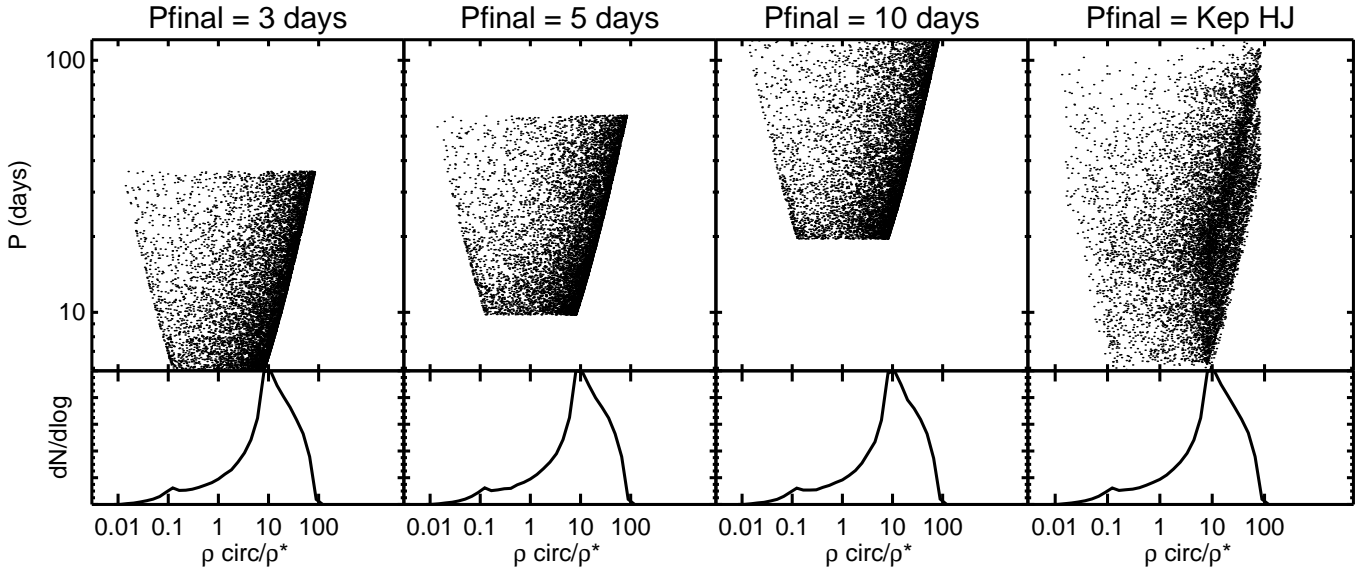}
\caption{Top: 2D posterior for orbital period $P$ vs. $\rhocirc/\rho_\star$ for planets with $0.6 < e < 0.9$ and $\pfinal = 3, 5, 10$~days (panels 1 - 3) or with $\pfinal$ drawn from \kep hot Jupiters in the interval $2.8 < P < 10$~days (panel 4). Bottom: Posterior $\rhocirc/\rho_\star$ marginalized over orbital period. Proto-hot Jupiters with $0.6 < e < 0.9$ do not typically have such large $\rhocirc/\rho_\star$ as their super-eccentric $(e > 0.9)$ counterparts (Figure~\ref{fig:twod}), making them less easy to identify. \label{fig:twotd}}
\end{figure*}

S12 focused their prediction on super-eccentric planets with $e > 0.9$. However, we also expect to find proto-hot Jupiters with less extreme eccentricities ($0.6 < e < 0.9$) along the same $\pfinal$ track. We repeat the procedure in \ref{subsec:super} for the interval $0.6 < e <0.9$. The overall occurrence rate for this interval is 0.61 relative to $\barNmod$. As shown in Figure~\ref{fig:twotd}, the proto-hot Jupiters in the $0.6 < e < 0.9$ range have shorter orbital periods ($6 < P < 121$~days). However, their transit durations and the inferred stellar density from a circular fit are not as strikingly anomalous as for the super-eccentric proto-hot Jupiters, making them less easy to identify. Therefore, we do not focus on these objects but discuss them further in the conclusion (Section 5). 

\section{Results: a Paucity of Proto-hot Jupiters}
\label{sec:obs}

We search for the super-eccentric proto-hot Jupiters predicted by S12 and find significantly fewer than expected. We describe our search procedure and present our measurements (Section \ref{subsec:search}) and assess the significance of this null result (Section \ref{subsec:signif}).

\subsection{Transit Light Curve Observables for Potential Proto-hot Jupiters}
\label{subsec:search}
We begin by identifying planet candidates that conform to our selection criteria. Applying the same criteria as in Section 2, we identify candidates with $8  R_\oplus < R_p < 22 R_\oplus$ and stellar parameters $4500 < \teff < 6500$~K and $\log g > 4$ (or, for those with $\log g < 4$, consistent with 4 within the uncertainty). We restrict the sample to candidates that have orbital periods between 34~days (corresponding to $\pfinal = 2.8$ for $e = 0.9$) and 2 years, transit three times in Q1-Q16, have signal-to-noise above 10, and which do not blatantly fail false-positive vetting. See Appendix \ref{app:selectlong} for further details. We are left with 31 planet candidates, including Kepler-419-b (D12). None of selection criteria disfavor eccentric candidates, nor does the pipeline penalize planets for having a duration different than that expected from a circular orbit. The consistency of our sample with Planet Hunters (discussed further in Appendix \ref{app:selectlong}), which were detected by eye, leads us to believe that the transits of long-period Jupiter-sized planets are not being missed due to transit timing variations. Moreover, Kepler-419-b -- our most eccentric planet --- was detected despite its 2 hour transit timing variations (TTVs).

For each candidate, we retrieve the publicly-available data from MAST. For each candidate, we use at least nine quarters of data (the amount available when we originally submitted this paper) and we supplemented with additional data for some candidates as we revised the paper. We extract the transits using {\tt AutoKep} \citep{photo2011G} and perform an MCMC fit using the Transit Analysis Package ({\tt TAP}; \citealt{photo2011G}). We fix $e=0$ but allow all other parameters to vary, including noise parameters for the \citet{photo2009C} wavelet likelihood function and first-order polynomial correction terms. We use short-cadence data when available. We obtain each candidate's $\rhocirc$ posterior. 

Next we follow\footnote{Instead of imposing a prior on the stellar mass, metallicity, and age from a TRILEGAL (TRIdimensional modeL of thE GALaxy; \citealt{k14742005G}) synthetic \kep field population, we assume a uniform prior on these model parameters, because a similar prior was already imposed by \citet{2013HS} to generate the stellar parameters.} Section 3.3 of D12 to compute a $\rho_\star$ posterior for each host star using the \citet{k14742007T} YREC stellar evolution models and the estimated effective temperature, surface gravity, and metallicity from \citet{2013HS}. We describe exceptions to this procedure in Appendix \ref{app:except}. 

\begin{figure}
\begin{centering}
\includegraphics{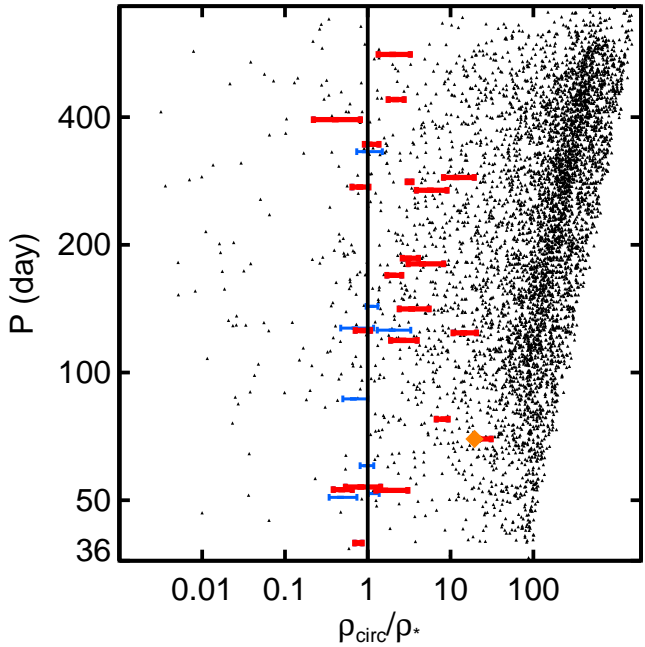}
\caption{Expected 2D posterior for orbital period $P$ vs. $\rhocirc/\rho_\star$ (taken from panel 4 of Figure~\ref{fig:twod}). The values we measured for our 34 candidates are overplotted.  Thin, blue bars: candidates with companions in their systems. Thick, red bars: candidates with no known companions. We do not see the expected half a dozen candidates in the region of high-probability density. Orange diamond: Kepler-419-b (bottom). \label{fig:koioverplot}}
\end{centering}
\end{figure}

Finally, we combine the $\rhocirc$ and $\rho_\star$ posteriors into a posterior of $\rhocirc/\rho_\star$ for each candidate, marginalized over all other parameters. In Figure~\ref{fig:koioverplot}, we plot the resulting values on top of the probability distribution for predicted super-eccentric proto-hot Jupiters (Figure~\ref{fig:twod}, panel 4). None of the candidates fall in the high-probability area of the prediction. We indicate candidates with known companions in their system with blue bars; none can have $e>0.9$ and $3<\pfinal<10$~days without its orbit crossing a companion's. As expected, all candidates with companions have $\rhocirc/\rho_\star$ close to 1.

Three candidates (none of which have known companions) have $\rhocirc/\rho_\star >10$: KOI-211.01, Kepler-419-b (D12), and KOI-3801.01 (first discovered by Planet Hunters, \citealt{2013W}, and now a \kep candidate). The probability of KOI-211.01 having $e > 0.9$ and $2.8<\pfinal<10$~days is 14\%.  D12 found that Kepler-419-b has $e = 0.81 ^{+0.10}_{-0.07}$ and $\pfinal = 14^{+6}_{-10}$~days. The probability of it having $e > 0.9$ and $2.8<\pfinal<10$~days is 12\%. Subsequently, $e>0.9$ was ruled out completely by our radial velocity follow up \citep{daw14}, in which we precisely measured $e=0.83\pm0.01$. Therefore we exclude this candidate from our statistical tests below. The probability of KOI-211.01 having $e > 0.9$ and $2.8<\pfinal<10$~days is 14\%. In assessing the consistency of the observations with the prediction of Section \ref{sec:expect}, we will fully consider the possibility that KOI-211.01 might be a super-eccentric proto-hot Jupiter.  However, we note that it is on the list of eclipsing binaries \citep{2011SP}. Finally, the probability of KOI-3801.01 having $e > 0.9$ and $2.8<\pfinal<10$~days is 8\%.

We emphasize that it is not surprising that many of the candidates lie in the low-probability region (likely Jupiters with $ e < 0.9$ or $\pfinal > 10$~days, of which there may be any number). It is only surprising that we do not see half a dozen in the high-probability region (given the assumptions in Appendix \ref{app:not}).

\subsection{Statistical Significance of Lack of Proto-hot Jupiters}
\label{subsec:signif}

None of the observed candidates lie in the high-probability-density region of Figure~\ref{fig:koioverplot}, so it is unlikely that the half a dozen predicted (Section \ref{subsec:expect}) super-eccentric proto-hot-Jupiters are present but missed. If we were certain that none of the candidates has $e > 0.9$ and $2.8<\pfinal<10$~days, the probability that observed number of super-eccentric proto-hot-Jupiters agrees with the prediction would simply be 1.3\%. (This is the probability, computed in Section 2, of observing 0 super-eccentric proto-hot Jupiters given the Poisson uncertainties in the observed number of super-eccentric Jupiters and in numbers used to compute the prediction.) However, there is a small chance that there are indeed super-eccentric proto-hot-Jupiters among the sample but that they just so happen to have their periapses oriented in the narrow range of angles producing an unremarkable $\rhocirc/\rho_\star$. Therefore we use a Monte Carlo procedure to assess the consistency of $\rhocirc/\rho_\star$ posterior derived for each candidate with the predicted population of super-eccentric planets. 

We first use the $\rhocirc/\rho_\star$ posteriors to generate an eccentricity posterior for each candidate, via a MCMC exploration of a limited set of parameters: $\rhocirc$, $\rho_\star$, $e$ and $\omega$ (as outlined in DJ12, Section 3.4). Although we can only make a tight eccentricity measurement when the planet's eccentricity is large (DJ12), the broad eccentricity posterior for the typical candidate here is useful for this purpose: it contains very little probability at the high eccentricities corresponding to $e > 0.9, 2.8<\pfinal<10$~days. We describe exceptions to this procedure in Appendix \ref{app:except}. Recently, \citet{2014K} (K14 hereafter) presented several caveats for deriving $e$ and $\omega$ from $\rhocirc$. We discuss these caveats and why they are not an issue for this study in Appendix \ref{app:star}.

We then perform $10^6$ trials in which we randomly select an eccentricity from each candidate's eccentricity posterior. We compute $\pfinal$ and count $\Nsup$ in Intervals 1 and 2. If both are greater than or equal to the respective numbers drawn from posteriors in Figure~\ref{fig:exp}, bottom panel (red dotted and blue dashed curves), we count the trial as a success, meaning that at least as many super-eccentric Jupiters as predicted were detected. 96.9\% of trials were unsuccessful. We exclude the candidates with known companions from this procedure (Figure \ref{fig:koioverplot}, thin blue bars), because it so happens that none of them can have $e>0.9$ and $2.8<\pfinal<10$~days without crossing the orbit of another candidate in the system. We find that, with 96.9\% confidence, we detected too few super-eccentric proto-hot Jupiters to be consistent with the prediction of Section \ref{sec:expect}. For example, 54\% of trials had 0 super-eccentric proto-hot Jupiters, 87\% had 1 or fewer, and 98\% had 2 or fewer. From these trials, we measure a $\Nsup$ posterior with a median $\Nsup = 0_{-0}^{+1}$ (89\% confidence interval). No single planet is likely to be supereccentric, but each has a small chance of being supereccentric.

The Jeffrey's prior we use in computing the posterior of the mean number of planets based on the observed number (Appendix \ref{app:poisson}) has a conservative influence on our results. If we were instead impose a uniform prior, the significance of the lack of supereccentric Jupiters would be 97.6\% instead of  96.9\%. The significance of our results would decrease if we had a prior expectation against either supereccentric Jupiters or the moderately eccentric Jupiters to which their occurrence is proportional, but we have no such expectation.

\section{Explaining the Paucity of Proto-hot Jupiters}
\label{sec:explain}

\begin{figure*}
\includegraphics[width=7in]{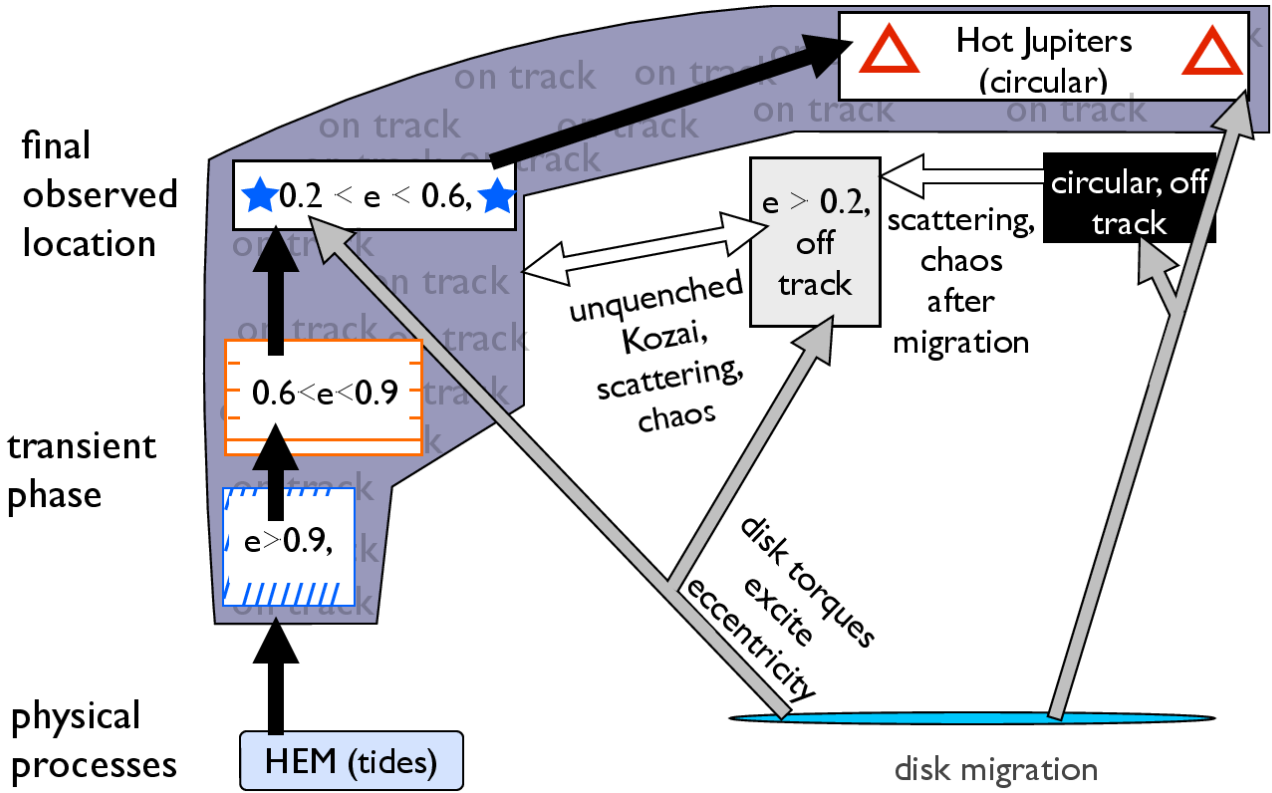}
\caption{ { Schematic of pathways (arrows) for creating the observed giant planet populations, which we assume form at orbital periods beyond about 2 years (corresponding to the uptick in giant planet frequency at around 1 AU, e.g. \citealt{metal2008C}) and migrated closer to their stars via HEM or disk migration. ``On track" (purple region) refers to the angular momentum range under consideration, i.e. $2.8<\pfinal<10$~days. The black arrows indicate the path that we have assumed throughout the paper for HEM caused either by a planetary or stellar perturber. For example, a Jupiter may be perturbed by a stellar binary companion, follow the black arrow to the region of super-eccentric Jupiters with $2.8<\pfinal<10$~days (blue, diagonal stripe region), undergo tidal circularization along its angular momentum track to $e < 0.9$ (horizontal orange striped region), become a moderately-eccentric Jupiter (blue stars), and eventually achieve hot-Jupiter-hood (red triangles). The other color arrows indicate alternative pathways caused by secular chaos, scattering, or unquenched Kozai moving the planets off track (white) or disk migration (gray), and the colors and patterns of the boxes correspond to the regions of parameter space in Figure~\ref{fig:obs}. See text for detailed discussion of the possible scenarios represented here.} \label{fig:diagram}}
\end{figure*}

So far (Sections 1Ñ3) we have been considering a scenario in which hot Jupiters begin beyond orbital periods of 2 years on super-eccentric orbits --- caused by gravitational perturbations from a companion (e.g. star-planet Kozai, planet-planet Kozai, planet-planet scattering, secular chaos) --- and subsequently undergo tidal circularization along a constant angular momentum track, reaching a final orbital period $\pfinal$. This process is known as high-eccentricity migration (HEM). We schematically summarize this (black arrows) and other possible origins for hot Jupiters (white and gray arrows), as well as moderately-eccentric Jupiters with $2.8<\pfinal<10$~days, in Figure~\ref{fig:diagram}. The corresponding populations from the non-\kep sample (EOD, \citealt{k14742012W}) are plotted in Figure~\ref{fig:obs}. 

Now we relax previous assumptions about HEM and explore how we can account for the lack of super-eccentric proto-hot Jupiters (Section \ref{sec:obs}). In Sections \ref{subsec:nohem} and \ref{subsec:bypass}, we relax the assumption that Jupiters began tidal circularization beyond an orbital period of two years, finding that this possibility could indeed account for the lack of super-eccentric Jupiters. In Section \ref{subsec:planetkozai}, we consider the particular case of HEM via the Kozai mechanism in which the Kozai oscillations of the observed moderately eccentric Jupiters are unquenched. In Section \ref{subsec:steady}, we relax the assumption of a steady current of hot Jupiters produced by HEM in the observed \kep sample but find that a lack of steady current is unlikely to account for the lack of super-eccentric proto-hot Jupiters. In Section \ref{subsec:synth}, we place an upper-limit on the fraction of hot Jupiters caused by Kozai perturbations from a stellar binary companion.  In Appendix \ref{app:not}, we describe additional assumptions, including tidal assumptions, most of which we do not expect to affect our results. 

\subsection{No Tidal Circularization: Hot Jupiters and Moderately-eccentric Jupiters Implanted Interior to 1 AU}
\label{subsec:nohem}

Rather than starting on highly-eccentric orbits with orbital periods above 2 years, hot Jupiters and moderately-eccentric Jupiters may have reached the region we observe today without tidal circularization. The moderately-eccentric Jupiters (blue stars, Figure~\ref{fig:obs} and \ref{fig:diagram}) observed along the angular momentum tracks have may have been placed there by whatever mechanism implanted eccentric Jupiters interior to 1 AU (gray region, \ref{fig:obs}). This population could possibly originate from disk migration. It remains debated whether planet-disk interactions could excite warm Jupiters' eccentricities. \citet{k14742003G} and \citet{k14742004S} argued that disk migration can potentially excite moderate eccentricities through resonance torques. Recently \citealt{metal2013DA} modeled planet-disk interactions using high-resolution three-dimensional simulations and found that disks are unlikely to excite the eccentricities of giant planets. However, \citet{tsang14} and \citet{tsang14b} made analytical arguments that planet eccentricities can be excited in a disks with entropy gradients. Planet-planet scattering directly from several AU is unlikely to be responsible for the moderately-eccentric Jupiters due to the large change in orbital energy required. Scattering following disk migration (e.g. \citealt{k14742011G}) is unlikely to produce sufficiently large eccentricities at such small semi-major axes because close encounters will lead to collisions rather than ejections \citep{ida13,petrovich14}. At such small semi-major axes, the relative velocities associated with a given eccentricity are large. Once giant planets stir each other to modest eccentricities, the relative velocities rival the escape velocities and the cross section for collision becomes a significant fraction of the scattering cross section. Collisions damp the relative velocities, limiting eccentricity growth. \citet{petrovich14} demonstrated the in-situ excitation cannot produce the observed moderate eccentricities of Jupiters within 0.15 AU (i.e. the blue stars in our Figure 1). However, secular chaos \citet{photo2011WL} following disk migration remains a possibility.

In Figure~\ref{fig:obs}, the moderately eccentric Jupiters (blue stars) look as if they could be an extension of the distribution of warm Jupiters in the gray region. In Table~\ref{tab:jupdense}, we compute the occurrence rate of giant planets $(M_p \sin i >0.25)$ in the RV-discovered sample in different regions of Figure~\ref{fig:obs}. The occurrence rate per log semi-major axis interval of moderately-eccentric Jupiters with $2.8<\pfinal<10$~days is less than or equal to that in the gray region ($10 < P < 250$~days). Therefore a separate mechanism for producing the blue stars apart from direct implantation may not be necessary. If non-tidal implantation was dominant, the number of moderately-eccentric Jupiters should not be used to predict the number of super-eccentric Jupiters because the moderately-eccentric Jupiters did not tidally circularize from super-eccentric orbits. 

\begin{deluxetable}{lllll}
\tabletypesize{\footnotesize}
\tablecaption{Occurrence\tablenotemark{a} of Jupiters detected by RV surveys (plotted in Figure \ref{fig:obs}).  \label{tab:jupdense}}
\tablewidth{0pt}
\tablehead{
\colhead{Period range}    & \colhead{Eccentricity}  & \colhead{Count} & \colhead{Poisson} & \colhead{Number}  \\
\colhead{(days)}    &  & & \colhead{range\tablenotemark{b}} & \colhead{per $\log_{10} a$}
}
\startdata
$2.8<\pfinal<10$& $0.2 < e < 0.6$ & 4 & $4 \pm 2$ & $11^{+6}_{-5}$ \\
$10 < P < 250$ & $0.2 < e < 0.6$ & 18&  $18^{+5}_{-4} $& $19^{+5}_{-4} $\\
\multicolumn{5}{c}{- - - - - - - - - - - - - - - - - - - - - - - - - - - - -- -  - - - - - - - - - - - -} \\
$2.8 < P < 5$ & $0< e < 0.2$ & 17 &$17 \pm 4$&$100^{+30}_{-20}$ \\
$2.8 < P < 10$& $0< e < 0.2$ & 21 &$21^{+5}_{-4} $ & $57^{+13}_{-11}$ \\
$5 < P < 10$ & $0< e < 0.2$ & 4 & $4.2^{+2.4}_{-1.7}$&$21^{+12}_{-9}$ \\
$5 < P < 250$& $0< e < 0.2$ & 29 &$29^{+6}_{-5}$ & $26^{+5}_{-4}$ \\
$10 < P < 250$& $0< e < 0.2$ & 25 &$25\pm5$ & $27^{+6}_{-5}$ \\
\enddata
\tablenotetext{a}{Numbers do not account for RV observational biases}
\tablenotetext{b}{Range of Poisson means from whose distributions the count could have been drawn, computed using Jeffrey's prior. See Appendix \ref{app:poisson} for further details.}
\end{deluxetable}

If the moderately-eccentric Jupiters with $2.8<\pfinal<10$~days did not undergo tidal circularization, hot Jupiters themselves could be part of a continuous distribution of circular Jupiters interior to 1 AU (Figure~\ref{fig:diagram} and \ref{fig:obs}, black region), which must have migrated somehow. Disk migration effectively produces planets on circular orbits, but seems inconsistent with the high obliquities of hot Jupiters orbiting hot stars (\citealt{k14742010W,metal2012A}, but see also \citealt{metal2012R,2013R}). However, disk migration may have produced some or all of the well-aligned hot Jupiters, if their low obliquities are not the result of tidal realignment.

If the cut-off for tidal circularization is 2.8~days, rather than 10~days, Jupiters on circular orbits with $P>2.8$~days would actually be part of the so-called ``period valley," rather than the hot Jupiter pile-up. The period valley refers to the region exterior to hot Jupiters with $P < 250$~days, where giant planets are scarce. The divide between hot Jupiters and the period valley (i.e. if it is 2.8~days, 10~days or some other value) is ambiguous in the literature  (e.g. \citealt{k14742003J,k14742003U,k14742009W,k14742010WO}). The observed ``edge" of hot Jupiters in ground-based transit surveys near 10~days may be partially caused by a combination of the reduced geometric transit probability of long-period planets and inefficiency of ground-based transit surveys in detecting them \citep{alias2005GSM}, rather than a drop in the intrinsic occurrence rate. We note that the distribution of giant planets inferred from the \kep Mission, assessed out to 50~days \citep{k14742011Y,k14742012HM}, has no such edge. However, \citet{k14742009W} detect an edge at approximately 0.07 AU (5~days) in their RV survey, which suffers from different (but less severe) biases than transit surveys. Thus, the existence and location of the cut-off remains uncertain.

The cut-off may in fact be \emph{between} $P = 2.8$~days and $P = 10$~days. In Figure~\ref{fig:obs}, the edge of the pile-up of circular Jupiters appears to end at around 0.057 AU (5~days), as \citet{k14742009W} found. If we separate the hot Jupiters below this cut-off, we recover a pile-up of hot Jupiters: in the region from $2.8 < P < 5$~days: we observe an excess of circular Jupiters inconsistent with the occurrence rate in the period valley by a factor of 3 (Table~\ref {tab:jupdense}). If the cut-off for hot-Jupiters is truly 5~days, the prediction for super-eccentric proto-hot Jupiters should be based only on the number moderately-eccentric Jupiters with $2.8 < \pfinal < 5$~days. In that case (repeating the calculations of Section \ref{subsec:expect}), we expect to find only $1\pm1$ super-eccentric proto-hot Jupiters with $e > 0.9$ and $2.8 < \pfinal < 5$~days (72\% confidence interval), and our confidence that we found fewer than predicted (Section \ref{subsec:signif}) drops to 72$\%$. 

\subsection{Some or All Proto-hot Jupiters May Have Bypassed the $e > 0.9$ Portion of the $\pfinal$ Track}
\label{subsec:bypass}

Alternatively, the typical hot Jupiter may have undergone tidal circularization but bypassed the high eccentricity phase, starting on the HEM track with $0.6 < e < 0.9$ in the region indicated by orange stripes in Figure~\ref{fig:obs} and \ref{fig:diagram} (or even in the $0.2 < e < 0.6$ region). For $\pfinal < 10$~days, a Jupiter would begin the HEM track at an orbital period less than 120~days, or 0.5 AU around a Sun-like star. The Jupiter is unlikely to have formed here --- the critical core mass required to accrete a massive atmosphere most likely exceeds the amount of refractory materials available \citep{metal2006R} (but see also \citealt{lee14}) --- or have been scattered here from several AU, which would require a large change in orbital energy. 

However, it may have been delivered to this region by disk migration followed by eccentricity excitation \citep{k14742011G} through secular chaos \citep{photo2011WL} or planet-planet scattering. However, in situ excitation by planet-planet scattering to the eccentricity values in the orange-striped region may not be feasible because the corresponding random velocity, $v_{\rm rand} \sim e \sqrt{GM_\star/a}$ is a significant fraction of a giant planet's escape velocity, $v_{\rm esc} = \sqrt{2 GM_p/R}$. In this super-Hill-velocity regime, as long as $v_{\rm rand} < v_{\rm esc}$, the cross section for collisions is of order $(v_{\rm rand}/v_{\rm esc})^2$ times the cross section for stirring (e.g. \citealt{gol04}).  For $v_{\rm rand} > v_{\rm esc}$, gravitational focusing is ineffective and only collisions occur.  Since $v_{\rm rand}/v_{\rm esc} \sim e (a/0.2 AU)^{-1/2} (M_p/M_J)^{-1/2}(R_p/R_J)^{1/2}$, $v_{\rm rand} \rightarrow v_{\rm esc}$ in the vicinity of the orange-striped region, and planet-planet interactions may be dominated by collisions rather than eccentricity growth. See \citealt{petrovich14} for an exploration of this phenomenon at smaller semi-major axes and \citet{ida13}, Figure 7 and Section 5.3 and \citet{beta2008J}, Figure 8 for scattering simulations in which the resulting eccentricities are lower for giant planets with $a < 1$ AU.

The overall picture of this scenario is that proto-hot Jupiters start the HEM track inside 1 AU with eccentricities similar to those of planets we observe in the period valley, rather than starting with $e \rightarrow 1$ at large semi-major axes. Assuming a steady-flux of proto-hot Jupiters into the orange striped region, if the four moderately-eccentric, RV-detected\footnote{In this calculation, we use the RV-detected sample. Even though transit probability is constant along an $\afinal$ track, ground-based transit survey are still strongly biased against detecting planets transiting with longer orbital periods. RV samples suffer from their own biases against long period and eccentric planets, which we do not account for here.}  Jupiters (corresponding to $\barNmod= 4\pm2$) in Figure~\ref{fig:obs} (blue stars) originated from the orange region, we would expect 
\begin{eqnarray}
\frac{\barNmod \int_{0.6}^{0.9} C_{\rm comp}(e) |\dot{e}|^{-1} de }{\int_{0.2}^{ 0.6 } C_{\rm comp}(e) |\dot{e}|^{-1} de }\nonumber \\
 = 4\pm2 \times 0.612 = 2.6^{+1.5}_{-1.0}\nonumber 
 \end{eqnarray} proto-hot Jupiters in the orange region. We see one such planet, HD~17156\,b  (Figure~\ref{fig:obs}); the probability of observing only one such planet is 5\%. Therefore we also observe a paucity of eccentric Jupiters in the orange region, so this is not a satisfactory explanation.

Since we observe HD~80606~b in the blue striped region, all proto-hot Jupiters must not begin in the orange striped region. Secular chaos or planet-planet scattering, taking place either where the planet formed or disk-migrated to, may place proto-hot Jupiters in both the orange striped region and the blue striped region, with the majority in orange striped region. The proto-hot Jupiters in these two regions would be created by the same dynamical processes responsible for Jupiters with $\pfinal > 10$~days (failed hot Jupiters), of which we observe more with $0.6 < e < 0.9$ than with $e > 0.9$ (though this may be partly due to observational bias).

\subsection{Proto-hot Jupiters Created by Planet-planet Kozai}
\label{subsec:planetkozai}

If a proto-hot Jupiter's Kozai oscillations have not been ``quenched" at the time of observation, we may observe the proto-hot Jupiter off its tidal-circularization track. The possibility that supereccentric proto-hot Jupiters spend time at low eccentricities does not reduce their expected number. S12 demonstrate that tidal dissipation primarily occurs during high eccentricity intervals. Regardless of how much time the Jupiter spends off its $\afinal$ track during low-eccentricity Kozai phases, it spends the same total amount of time on the $\afinal$ track undergoing tidal dissipation. Therefore Equation (\ref{eqn:predn}) predicts the total number of super-eccentric Jupiters \emph{observed on the track}. 

However, it is possible that the Kozai oscillations of the moderately-eccentric calibration proto-hot Jupiters ($\Nmodz$ blue stars in Figure \ref{fig:obs}) are not quenched.  We may be observing them in the low-eccentricity portions of their cycles, and they may actually be tidally dissipating on a track with $\pfinal < 2.8$ days. If so, they should not have been used to compute the expected number of super-eccentric proto-hot Jupiters with $2.8 < \pfinal < 10$ days. We clarify that even if the moderately eccentric Jupiters oscillate to high eccentricity intervals corresponding to a dissipation track with $\pfinal < 2.8$ days, they should not be used to compute the number of super-eccentric Jupiters with $\pfinal < 2.8$ days, because Equation (\ref{eqn:predn}) only applies to planets observed on the track. Supporting this possibility, while hot Jupiters themselves do not have nearby companions \citep{metal2012SR}, \citet{2014D} find that of the sample of warm Jupiters with known companions, all the companions are sufficiently massive and nearby for the Kozai cycles to remain unquenched. Additionally, several of the moderately-eccentric Jupiters (HAT-P-34~b, HAT-P-31~b, and WASP-8~b) have linear trends in the RV observations \citep{photo2011W}, indicating the presence of a companion.

We rewrite the condition for Kozai cycles to not be quenched from S12 (Equation 12) as (assuming a sun-like star and circular perturber with a minimum mutual inclination near $40^\circ$):
\begin{equation}
a_{\rm per} \le 15 {\rm AU} \left( \frac{1-0.32^2}{1-e^2} \right) \left(\frac{\afinal}{0.05 \rm AU}\right)^{8/7} \left(\frac{M_{\rm per}}{M_\odot}\right)^{2/7}
\end{equation}
\noindent for which $a_{\rm per}$ and $M_{\rm per}$ are the semi-major axis and mass of the putative perturbing companion, and $\afinal$ and $e$ are the quantities for the moderately eccentric proto-hot Jupiter at the peak eccentricity of its Kozai cycle. All of the warm Jupiters in our sample have been observed via RV observations and such a nearby star would be evident via a dual set of lines or major inconsistency in the RV forward-modeling process. Therefore the perturber must be a planet, so hereafter we will refer to the possibility discussed in this section as planet-planet Kozai.

\subsection{Alternatives to the ``Steady Current" Approximation}
\label{subsec:steady}

The S12 prediction of a readily observable number of super-eccentric proto-hot Jupiters assumed a ``steady current" of proto-hot Jupiter production based on the fact that stars have been steadily produced throughout the Galaxy. In a sample of stars of identical ages, we would expect a steady current of proto-hot Jupiters only if the rate of hot Jupiter production throughout a star's lifetime is constant (e.g. that a hot Jupiter is just as likely to be produced between 4.1-4.2 Gyr as it is during the first 100 Myr). This seems unlikely. In the HEM mechanisms proposed (Section 1), proto-hot Jupiters are spawned (i.e. begin their HEM journey) on instability timescales (planet-planet scattering, secular chaos) or the Kozai timescale, which are unlikely to always coincide with the typical stellar lifetime. More likely, the distribution of timescales is uniform (or normal) in order of magnitude and thus most proto-hot Jupiters are spawned early in their host stars' lifetimes. Indeed, \citet{photo2012Q} recently discovered hot Jupiters in the 600 Myr Beehive cluster and found that, accounting for the cluster's enhanced metallicity, the hot Jupiter occurrence rate is consistent with that of the solar neighborhood.

In a sample of stars with uniform random ages spanning 0 to the age of the Galaxy, we would expect a steady current, even if most proto-hot Jupiters are spawned early in their host stars' lifetime, as long as the conditions for forming hot Jupiters have not changed over time. In practice we expect young stars to be rotating too rapidly to be amenable to Doppler observations and too uncommon in our stellar neighborhood to make up a representative sample in transit surveys. Therefore, because of selection bias, the steady current approximation is unlikely to exactly hold.

\begin{figure*}
\includegraphics[width=\textwidth]{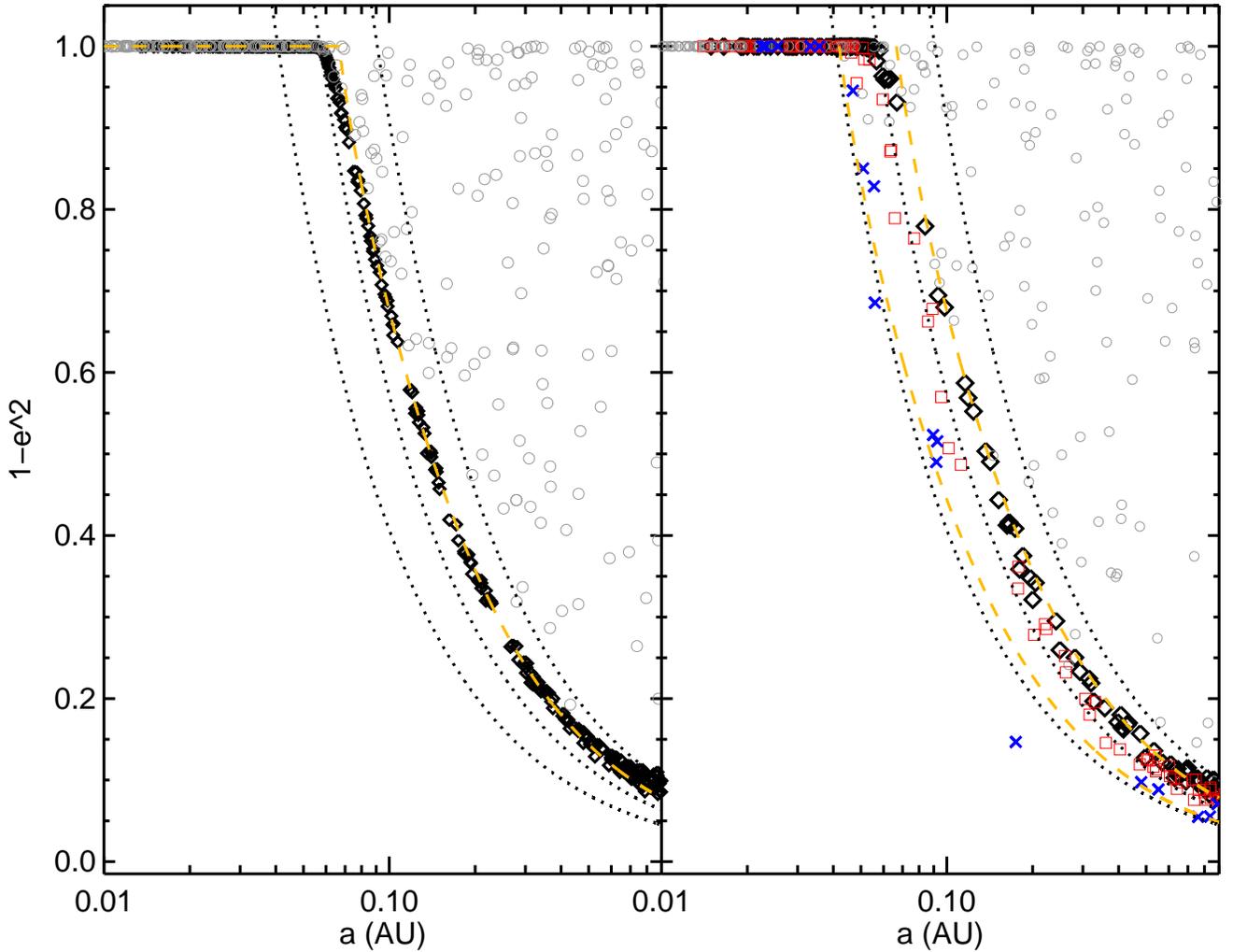}
\caption{Monte Carlo simulation of tidal evolution of proto-hot Jupiters assuming that all the proto-hot Jupiters were created at once (left panel) or that there is a steady current (right panel). In each simulation, proto-hot Jupiters are drawn from a distribution uniform in eccentricity and semi-major axis, extending to 10 AU. Planets that began interior to 1 AU are marked as open, gray circles.  We overplot tracks of constant angular momentum (dotted lines) corresponding to $\pfinal = 3, 5, 10$~days, as well as lines defined by a constant  ``orbital change timescale" (dashed orange lines, Equation \ref{eqn:move}). In the right panel, the outer orange-dashed line represents a timescale 40 times longer than the inner orange-dashed line. The red squares (blue x) were created two (thirteen) times more recently than the black diamonds. \label{fig:mcmc}}
\end{figure*}

However, even if all the proto-hot Jupiters in our sample were spawned simultaneously, \emph{we would still expect to observe proto-hot Jupiters.} However, they would be restricted to the narrow range of $\afinal$ tracks for which the circularization timescale is of order a star's age, instead of being found along all $\afinal$ tracks in proportion to the circularization timescale ($e/\dot{e} \propto \afinal^8$). Inspired by population simulations by \citet{photo2010H} and \citet{2012H}, we simulate an extreme scenario in which every proto-hot Jupiter in the observable sample is created simultaneously (Figure~\ref{fig:mcmc}, left panel). We begin with a population of Jupiters uniformly distributed in eccentricity and semi-major axis, extending to 10 AU; gray open circles had initial semi-major axes interior to 1 AU (representing the possibility that planets can begin HEM interior to 1 AU, as discussed in Section \ref{subsec:bypass}). In reality, the initial conditions will be set by whatever dynamical processes excite the planet's eccentricity from the circular orbit it formed on. Since here we are agnostic about which of the proposed dynamical mechanisms is at play for initially exciting the proto-hot Jupiter's eccentricity, we use a uniform distribution to get a sense for the evolution.

Then we evolve the tidal evolution equations until Jupiters with $\pfinal < 5 $~days have circularized. This is not absolute timescale but relative to the unknown tidal dissipation constant. In Figure 7, we overplot tracks of constant angular momentum (dotted lines), as well as lines defined by a constant ``orbital change timescale" set equal to the time of the plotted snapshot,

\begin{equation}
\label{eqn:move}
t_{\rm move}~=~\left[\left(\dot{a}/a\right)^2+\left(\dot{e}/e\right)^2\right]^{-1/2},
\end{equation}
(dashed orange lines) which match the $\afinal$ tracks at low eccentricities. This relationship represents an envelope in $(a,e)$ space for a population of planets undergoing migration over a particular timescale $(t_{\rm move})$, rather than the evolution of a particular planet along a track.  Although there is no steady current, we see a ``track" consisting of a) Jupiters along the same $\afinal$ track but with different starting eccentricities/semi-major axes, and b) Jupiters along close, adjacent $\afinal$ tracks (those along the slightly larger $\afinal$ track have higher eccentricities because $e/\dot{e} \propto \afinal^8$). 

For comparison (Figure~\ref{fig:mcmc}, right panel) we perform a simulation featuring a steady current of hot Jupiters. In this case, the proto-hot Jupiters are distributed over a range of angular momenta tracks but are most common (black diamonds) along the track where the tidal circularization time is order the total elapsed time (i.e. most of the proto-hot Jupiters began their HEM early in the lifetime of the oldest stars in the sample). The smaller $\afinal$ tracks (red squares, blue X) are more sparsely populated because these planets circularize very quickly and we just happen to be catching some. The left and right panels are not strikingly different. Particularly with a small observational sample size, we are unlikely to be able to be able to distinguish whether we are seeing a narrow range of $\afinal$ tracks due to a lack of steady current (left) or simply due to a higher relative population along the $\afinal$ track of order a stellar age (right).

However, without a steady current of hot Jupiters, the number of moderately-eccentric Jupiters in a $\pfinal$ range cannot be used in Equation~(\ref{eqn:rC}) to predict the number of super-eccentric Jupiters. The number of super-eccentric Jupiters would depend on the initial conditions generated by HEM mechanisms: the relative number of hot Jupiters along adjacent angular momentum tracks and beginning at different semi-major axes along the same track. However, if the initial eccentricities are roughly independent of semi-major axis, the distribution today would resemble one resulting from a steady current. The observed lack of super-eccentric proto-hot Jupiters would require very fine-tuned initial conditions, such as substantially fewer Jupiters beginning along a slightly larger angular momentum track. Therefore the paucity we found in Section 3 is unlikely to be fully accounted for by a lack of steady current.

Throughout this work, we have assumed that proto-hot Jupiters \emph{can} travel along HEM tracks with circularization timescales of order a stellar lifetime. However, a nearby planetary perturber can potentially permanently remove the proto-hot Jupiter from the angular momentum track before tidal circularization decouples it. For example, in secular chaos, a planet may be perturbed to a high eccentricity, begin to circularize along a track with a timescale longer than the chaos timescale, but then be chaotically perturbed by nearby planets again.  If all proto-hot Jupiters are created early in the star's lifetime and only those with extremely short tidal circularization timescales escape the perturbations of nearby planets, we would indeed see a lack of super-eccentric proto-hot Jupiters in a sample that lacks young stars. In this framework, the observed moderately-eccentric Jupiters would not have been produced by HEM but by some other mechanism, as explored in Section \ref{subsec:nohem}. Their survival indicates that the circularization timescale is not extremely short in $2.8<\pfinal<10$~days, and therefore the argument that planets can only travel along paths with very short circularization timescales would only apply for hot Jupiters with $P < 2.8$~days.

Finally, a related issue is whether the ratio of moderately-eccentric Jupiters to Jupiters with $P=\pfinal$ is the same in the \kep and calibration samples (i.e. $\barNmod/\barNpp \simeq \barNmodz/\barNppz$), as assumed in Equation~(\ref{eqn:nmod}). We would expect the pile-up of hot Jupiters to be greater in the older sample (i.e. the \kep sample) and therefore that Equation  (\ref{eqn:nmod}) might over predict the number of super-eccentric Jupiters. However, we note that the overall occurrence rate of hot Jupiters in the \kep sample is actually \emph{smaller} than the in the RV-sample \citep{k14742012HM,k14742012W}, so this is unlikely to be a problem in practice.

\subsection{Upper Limit on Star-planet Kozai Contribution}
\label{subsec:synth}
If the observed moderately-eccentric proto-hot Jupiters began with orbital periods above 2 years and with $ e \rightarrow 1$ and underwent tidal circularization while staying on a track of constant angular momentum (black arrows, Figure~\ref{fig:diagram}), we would expect to see four super-eccentric proto-hot Jupiters (Section 2); the lack of such planets indicates that one or more of the alternative pathways in  Figure~\ref{fig:diagram} (white and gray arrows) may dominate. These alternative pathways all originate from a planetary perturber or disk, rather than a stellar perturber (see Section \ref{subsec:planetkozai} for a justification of why a planetary Kozai perturber would be required). Here we place an upper limit on the fraction of hot Jupiters that followed the black arrow channel of HEM, beginning with a super eccentricity and moving along a track of constant angular momentum. Since this is the only pathway open to hot Jupiters produced by star-planet Kozai oscillations, the upper limit is also on the fraction of hot Jupiters created by stellar binaries. We repeat the MCMC procedure in Section \ref{subsec:signif} but update Equation~(\ref{eqn:predn}) with an additional parameter, $f_\star$, representing the fraction of hot Jupiters that undergo HEM from super-eccentricities (or, equivalently, the maximum fraction produced by star-planet Kozai):

\begin{equation}
\label{eqn:prednr}
\barNsup = f_\star r(\emax) \barNmod.
\end{equation}

We impose a modified Jeffrey's prior on $f_\star$, so the prior is uniform between 0 and 5\% and scales with $1/f_\star$ above $f_\star$ = 5\%. We obtain a two-sigma limit of 44\% on the fractional contribution from stellar binaries. Therefore, we expect at least half of hot Jupiters were created by a planetary perturber (or a disk). We note that this result technically only applies to hot Jupiters with $P > 2.8$~days, because no super-eccentric Jupiters with $\pfinal < 2.8$~days were expected. This limit is consistent with 30$\%$ contribution rate found by \citet{photo2012N}. This limit also implies that we would need at least a 44$\%$ false positive rate to account for the discrepancy. As discussed in Appendix \ref{app:not}, such a high false positive rate is unlikely based on previous studies, and moreover, our additional vetting (Appendix \ref{app:selectlong}) results in our sample containing fewer false positives than the \kep candidate list.

\section{Conclusion}
\label{sec:conclude}
S12 predicted that if high-eccentricity migration (HEM) is the primary channel for producing hot Jupiters, the \kep candidate collection should harbor a population of super-eccentric Jupiter-sized planets that are in the midst of tidal circularization. We developed and implemented a procedure to use the publicly-available \kep transit light curves to test this prediction and found a paucity of proto-hot Jupiters on super-eccentric orbits. Incorporating uncertainties due to counting statistics, uncertainties in the fitted light curve parameters and in the assumed stellar parameters, incompleteness due to the limited observational timespan and missing data, and the signal-to-noise limit, we expected to observe $\Nsup = 4$ (Section \ref{sec:expect}) but instead found only $0_{-0}^{+1}$  (Section \ref{sec:obs}). (Note that we did not find one Jupiter that is likely to be supereccentric; rather, each Jupiter has a small chance of being supereccentric but with its orbit oriented so that its transit speed is unremarkable.) False positives are unlikely to account for the discrepancy. The lack of super-eccentric proto-hot Jupiters may indicate that the assumed constant tidal time lag approximation --- which sets the ratio of super-eccentric proto-hot Jupiters to the observed, partially circularized moderately-eccentric Jupiters used to compute the prediction --- is incorrect (Appendix \ref{app:not}). However, violation of this assumption could only account for the discrepancy if tidal dissipation were actually much \emph{stronger} at high eccentricities along a given angular momentum track. The statistical significance of our results is 96.9\%. In the future, we will use a pipeline targeted to finding large-period, Jupiter-sized planets (including those with only two transits, allowing us to extend the maximum orbital period considered from 2 years to 4 years) and inject the transits of simulated super-eccentric Jupiters to ensure that the assumptions about completeness we have made here are correct.

In Section \ref{sec:explain}, we explored a number of dynamical explanations for the paucity of super-eccentric proto-hot Jupiters, relaxing the assumptions by S12 that proto-hot Jupiters begin HEM beyond a 2 year orbital period and that a steady current of hot Jupiters is being produced in the observed \kep sample. We found that the lack of super-eccentric planets could be explained by one of the following scenarios. First, hot Jupiters with $P > 2.8$~days may be implanted interior to 1 AU, and only those with $P< 2.8$~days have undergone tidal circularization. This would be the case either if the tidal circularization timescale is typically only less than a stellar lifetime for $\pfinal < 2.8$~days or if only proto-hot Jupiters with fast circularization timescales can manage to complete their circularization without being moved by a nearby planetary perturber. Second, the moderately-eccentric Jupiters used to calibrate the prediction may be undergoing Kozai eccentricity oscillations caused by a nearby planetary perturber and we are observing them in the low-eccentricity phase, in which they are not currently undergoing tidal dissipation. In that case, we would not expect to observe any super-eccentric Jupiters currently undergoing tidal dissipation.

All these explanations point either to disk migration or to secular chaos, planet-planet scattering, or planet-planet Kozai (or other yet-to-be-proposed dynamical mechanism) as the dominant channel for hot-Jupiter production, rather than the star-planet Kozai mechanism. In Section 4, we placed an upper limit of 44\% on the contribution of star-planet Kozai to hot Jupiters, consistent with the findings of \citet{photo2012N}. Our limit only applies to hot Jupiters with orbital periods greater than 2.8~days, as the prediction for super-eccentric Jupiters only applied to those ending their HEM journey at $2.8 < P < 10$~days.

In this paper, we explored S12's prediction for proto-hot Jupiters, but they made a similar prediction for a population super-eccentric binary stars, which they subsequently discovered \citep{photo2012D}. It would not be surprising if short-period stars were produced by the Kozai mechanism but short-period planets primarily by scattering and chaos, which can potentially deliver the planets observed interior to 1 AU without the planets undergoing a super-eccentric phase. The initial conditions for stellar systems and planetary systems may differ in that planetary systems are both theorized (e.g. \citealt{photo2004B}) and observed (e.g. \citealt{k14742009W,photo2011L,k14742012MM}) to form packed with many planets, a condition that may often lead to scattering and secular chaos. In contrast, stellar multiples are typically hierarchical, an optimal setup for the Kozai mechanism.

The lack of super-eccentric proto-hot Jupiters is a new piece of evidence that models for the origins of hot Jupiters' close-in orbits must match, joining the distribution of spin orbit measurements (e.g. \citealt{nep2009F,photo2011MJa,photo2012N}). We recommend that future theoretical studies of dynamical models for forming hot Jupiters predict: the distribution of timescales for instabilities that form proto-hot Jupiters, how often the high-eccentricity phase of HEM is bypassed, the initial conditions along the HEM angular momentum tracks, and the expectations for high-eccentricity ``failed'' hot Jupiters that likely have periapses too distant to undergo tidal circularization, such as Kepler-419-b (D12).  For the brightest \kep host stars, we recommend measuring the spin-orbit alignment of planets in the period valley, whose obliquities have presumably not been affected by tides. Such measurements could elucidate whether the planets in the period valley have a single origin or if there are two populations, which might correspond to the circular planets and the eccentric planets. Additionally, we recommend investigating whether a gas disk could flatten and circularize a period valley planet's orbit if the planet were scattered there before the gas disk dissipated.

We recommend that observers strive to better characterize the eccentricity distribution of the period valley, which we argued may be the launching point for the typical hot Jupiter's HEM journey. It would be helpful to assess if the occurrence rate of eccentric Jupiters in this region is -- when extrapolated to the $3<\pfinal<10$~days region -- sufficient to launch all the hot Jupiters interior to 1 AU. We also recommend that observers attempt to pin down the period or semi-major axis cut-off between the hot Jupiter pile-up and the period valley. Finally, although we found that it would be more difficult to identify proto-hot Jupiters with $0.6 <e < 0.9$ using the ``photoeccentric effect,'' it could be feasible with more accurate and precise stellar parameters. We recommend spectroscopic follow-up of KOI host stars for this purpose.

\acknowledgments
We are thankful to the anonymous referee(s) for an insightful report that improved the paper and in particular for advocating more conservative assumptions about the completeness of the \kep pipeline. R.I.D. gratefully acknowledges the National Science Foundation Graduate Research Fellowship under grant DGE-1144152 and the Miller Institute for Basic Research in Science, University of California Berkeley. J.A.J. is grateful for the generous grant support provided by the Alfred P. Sloan and David \& Lucile Packard foundations. We are grateful to Smadar Naoz for many enlightening discussions and comments, including opening our eyes to other possibilities in Section 4, for which we also thank Simon Albrecht and Fred Rasio. Many thanks to Adrian Barker, Rick Greenberg, Renu Malhotra, Francesca Valsecchi, and especially Brad Hansen for tidal insights; to Katherine Deck, Will Farr, Vicky Kalogera, Yoram Lithwick, and Matthew Payne for helpful dynamical discussions; to Will Farr and Moritz G\"{u}nther for helpful statistical discussions; to Courtney Dressing and Francois Fressin for \kep assistance; and to Joel Hartmann for helpful discussions about the ground-based sample. We are grateful to Subo Dong for constructive comments on a manuscript draft. Thanks to Joshua Carter, Boas Katz, Doug Lin, Geoff Marcy, Darin Ragozzine, Kevin Schlaufman, and Aristotle Socrates for helpful comments and to Thomas Barclay, Christophe Burke, Jon Jenkins, and Jason Rowe for helpful discussions of \kep's completeness for giant planets. We benefitted from helpful conversations with Thomas Barclay about candidate vetting and Howard Isaacson and Rea Kolb about false positives. We are very grateful to Chelsea Huang for helpful discussions and for providing us with a detrended light curve for KIC~6805414. R.I.D. thanks David Charbonneau, Sean Andrews, Debora Fischer, Matt Holman, and Abraham Loeb for helpful comments on the thesis chapter version of this manuscript.  Special thanks to J. Zachary Gazak for helpful modifications to the TAP code. We thank D\'{a}ith\'{i} Stone for making available his library of IDL routines.

This paper includes data collected by the \kep mission. Funding for the \kep mission is provided by the NASA Science Mission directorate. We are grateful to \kep team for all their work in making this revolutionizing mission possible and making available the rich \kep dataset. Some of the data presented in this paper were obtained from the Multimission Archive at the Space Telescope Science Institute (MAST). STScI is operated by the Association of Universities for Research in Astronomy, Inc., under NASA contract NAS5-26555. Support for MAST for non-HST data is provided by the NASA Office of Space Science via grant NNX09AF08G and by other grants and contracts. This research has made use of the Exoplanet Orbit Database. This research has made use of the NASA Exoplanet Archive, which is operated by the California Institute of Technology, under contract with the National Aeronautics and Space Administration under the Exoplanet Exploration Program.

\appendix

\section{Sample of Long-Period \kep Giant Planet Candidates}
\label{app:selectlong}
We check for consistency among several sources to ensure that our sample is complete for planets that transit three times in Q1-Q12:
\begin{itemize}
\item The \citet{k14742013B} candidate sample, which is complete for Jupiter-sized planets that transit three times within Q1-Q6, and also contains a number of planets that transit one or more times in Q1-Q8, with unknown completeness.
\item The Q1-Q12 threshold-crossing events table (TCE; acquired from the NASA Exoplanet Archive; see \citealt{2012T,2013T}), which is complete for Jupiter-sized planets that transit three times within Q1-Q12 ($\tsurvey$ = 3 years), but the objects have been not necessarily been vetted as candidates. We compile of a list of TCE that meet our stellar and planetary criteria and remove those that are obviously instrument or stellar variability artifacts or exhibit a deep secondary eclipse (indicating that they are eclipsing binaries). We use the data validation reports available for each TCE at the Exoplanet Archive to also exclude TCE with different odd and even transit depths (indicating that they are eclipsing binaries) or centroids that strongly correlate with the transits, indicating a blend. In total we find four candidates not present in the \citet{k14742013B} sample:  KIC 12735740 (orbital period 282 days, ow KOI-3663.01), KIC 8827930 (orbital period 288 days, now KOI-3801.01), KIC 8813698 (orbital period 269 days, ow KOI-1268.01), and KIC 9025971 (orbital period 141 days, now KOI-3680.01). The first three exhibited two transits in Q1-Q8 but were apparently not caught by eye. The fourth did not exhibit two transits, despite its shorter orbital period, due to missing data.
\item The \citet{metal2013BB} sample, which is complete for Jupiter-sized planets that transit three times within Q1-Q8. KIC 8813698 (above) is now included as KOI-1268.01. 
\item The Q1-Q12 sample at the Exoplanet Archive. This sample now includes all four candidates we found on the TCE table: KIC 8813698 (now KOI-1268.01), KIC 8827930 (now KOI-3801.01), KIC 9025971 (now KOI-3680.01), and KIC 12735740 (now KOI-3663.01).
\item The Q1-Q16 sample at the Exoplanet Archive, which is complete for Jupiter-sized planets that transit three times within Q1-Q16 ($\tsurvey$ = 4 years). 
\item The catalogue of long-period planets discovered by Planet Hunters \citep{2013W,2013S}. Of the 42 discoveries reported in \citet{2013S}, three correspond to those we identified from the TCE table (KIC 8827930, 9025971, 12735740). KIC 9413313 transits three times in Q1-Q16; its strong stellar variability may be the reason it did not appear on the Q1-Q16 candidate or TCE list. We add this candidate to our sample. Twenty-eight candidates fell below our planetary radius cut (KIC 2975770, 3326377, 3634051, 3732035, 4142847, 4472818, 48rd50, 4902202, 4947556, 5857656, 5871985, 5966810, 7826659, 8210018, 8636333, 9147029, 9166700, 9480535, 9886255, 10024862, 10360722, 10525077, 10850327, 11026582, 11253827.01, 11253827.02, 11392618, 11716643). KIC 4760478b and the planet candidate orbiting 9663113 do not transit three times in Q1-Q16. KIC 8012732 does not have KIC stellar parameters available. For six others (KIC 3663173, 5437945.01, 5437945.02, 6878240, 9425139, 9958387), when we fit the reported stellar parameters with the \citet{k14742007T} models, we obtained stellar radii much smaller than those reported by \citep{2013W}, resulting in a planetary radius below our cut. Of the fourteen discoveries reported by \citet{2013S}, four stars fall outside our stellar cuts (2437209, 5010054, 5522786,  6805414), and eight have planets falling above or below our planetary radius cuts (5094412, 6436029, 9662267, 9704149, 10255705, 11152511, 11442793, 12454613). KIC 5732155b does not transit three times in Q1-Q16. KIC 6372194b appears on the Q1-Q16 candidates list but with a planetary radius below our cut, due to the smaller planet-to-star radius ratio than that reported by \citet{2013S}. We fit the light curves using the approach described in Section \ref{subsec:search} and also found a small planet-to-radius ratio ($R_p/R_\star = 0.085 \pm 0.003$), resulting in $R_p < 8 R_{\rm earth}$. Therefore we do not include this candidate in our sample.
\item  The catalogue by \citet{2013H} discovered using the HAT pipeline. Of the new candidates discovered by \citet{2013H}, only KIC 6805414 falls within our stellar and planetary cuts (KIC 5563300 falls below our planetary radius cut using the updated stellar parameters at the Exoplanet Archive). KIC 6805414 appears on the Q1-Q16 candidate list (KOI-5329.01) and may have been missed in Q1-Q12 by the \kep pipeline because strong stellar variability occurs on a timescale similar to the transit duration. However, KIC 6805414 falls below our surface gravity cut using the updated parameters by \citet{2013S}.
\item The catalogues of false positives (acquired from the NASA Exoplanet Archive) and eclipsing binaries (acquired from MAST). However, of the objects meeting our planetary radius cut, all had obvious secondary eclipses, even-odd eclipse depths, or centroid shifts. KOI-211.01 is on the list of \kep eclipsing binaries but may be a planet so we keep it in our sample. 
\item A catalogue of long-period candidates created by an amateur astronomer \citep{2012P}. The \citet{2012P} catalogue contains four candidates not in the \citet{metal2013BR} sample, three of which we identified above (KIC 12735740 , KIC 8827930, and KIC 9025971) and one of which is the planet candidate identified by \citet{2013H}, orbiting KIC 6805414.
\end{itemize}
Because these sources are consistent when compared across which quarters they probed, we have confidence that the giant planet candidates transiting three times are generally being picked up by the pipeline. 

Our final list of long-period candidates is: KOI-209.01*, KOI-211.01, KOI-351.01*, KOI-372.01, KOI-398.01*, KOI-458.01, KOI-682.01, KOI-806.01*, KOI-806.02*, KOI-815.01, KOI-918.0, KOI-1089.01*, KOI-1193.01, KOI-1268.01, KOI-1353.01*, KOI-1431.01, KOI-1439.01, KOI-1466.01, Kepler-419-b, KOI-1477.01, KOI-1483.01, KOI-1486.01*, KOI-1552.01, KOI-1553.01, KOI-1587.01, KOI-2689.01, KOI-3680.01, KOI-3801.01, KOI-5071.01, KOI-5241.01, and KIC 9413313. Those with asterisks are in systems with multiple candidates. 

Now we list a subset of candidates that appear on candidates list but that we exclude from our sample and describe our motivation for excluding them. The main reason why we end up excluding so many candidates is that the current \kep catalogue is very generous in including candidates that may be false positives, and the vetting of the current catalogue is not uniform. We apply uniform vetting criteria to both the long-period planets and the hot Jupiters, making use of the data validation report.

 We exclude the following candidates because they exhibit secondary eclipses inconsistent with planethood: KOI-193.01, KOI-856.01 \citep{2013O}, and KOI-3641.01. We exclude KOI-433.02 because it has large centroid-transit correlations (including an out-of-transit offset of 0.3 arcseconds), as well as KOI-617.01 (0.7 arcseconds), KOI-772.01 (0.2 arcseconds), KOI-1645.01 (0.3 arcsecond), KOI-1773.01 (0.3 arcseconds) KOI-3678.01 (0.4 arcsecond),   KOI-3709.01 (0.7 arcsecond), KOI-3717.01 (0.5 arcsecond), and KOI-3726.01 (0.8 arcsecond, as well as different odd/even eclipse depths). We also exclude KOI-620.02, KOI-625.01, KOI-1137.01, KOI-1242.01, KOI-1255.01 (which \citealt{2013M} find also shows transit and rotation modulation on different stars), KOI-1684.01, KOI-1691.01, KOI-855.01, KOI-3660.01, KOI-4939.01, KOI-5312.01, KOI-5385.01, KOI-5488.01, and KOI-5792.01 due to strong centroid-flux correlations, which are both statistically significant and strikingly evident by eye as a strong ``wind'' in flux vs. position ``rain'' plot. We exclude KOI-1095.01 because it has large centroid (0.1 arcseconds) and source (0.2 arcseconds) offsets and a noticeable flux-centroid correlation; moreover, its radius is below 8 Earth radii in the Q1-Q12 and Q1-Q16 candidates list. KOI-366.01, KOI-377.02, KOI-777.01, KOI-1783.01, and KOI-1787.01 appeared on earlier candidates list but no longer meet our stellar and/or planetary criteria due to revised stellar properties. We do not include KOI-1288.01 because of revised stellar parameters (see Appendix \ref{app:except}).  KOI-1335.01 had a radius above 8 Earth radii in the Q1-Q8 candidate list, but in the Q1-Q12 candidate list, Q1-Q16 data validation report, and our own fits, it has a radius below 8 Earth radii, so we do not include it here. Candidate KOI-1496.01 is too small, and stars KOI-2659, KOI-5314 , and KOI-5576 are too giant. (KOI-2569 was identified as a giant by \citealt{2013M}). KOI-375.01, KOI-422.01, and KOI-490.02 only transit twice in Q1-Q16; KOI-1096.01 only transit once. We do not include KOI-686.01, which we analyzed in detail in DJ12, because it was found by Diaz et al. (2013, in preparation) to be a brown dwarf. KOI-771.01 and KOI-3787.01 have different odd-even eclipse depths and are likely false positives, so we do not include them here; KOI-5018.01 and KOI-5760.01 also have inconsistent depths. We also discard KOI-3320.01, which was originally on the eclipsing binaries list due to its large radius, because its position is highly offset from the KIC catalogue, implying that the star is probably misidentified and therefore mischaracterized. Star KOI-3678 lacks KIC parameters. KOI-5409.01 only has one quarter of data and KOI-5682.01 only two. KOI-5446.01, KOI-5697.01, and KOI-5923.01 have low SNR (10, 9.4, and 7 respectively). KOI-5661.01 appears to be an instrumental artifact and KOI-5802.01 an artifact of stellar variability.

Several candidates that would cross the orbits of other candidates in their systems if they were highly-eccentric also have very grazing transits that are difficult to fit (KOI-734.02, KOI-1258.03, KOI-1426.03) or are short period with many transits to extract (KOI-620.01); therefore we do not bother to fit them. We exclude KOI-1356.01 (which does not transit three times in Q1-Q12) because its true period is 787.45 days, far outside the completeness range we consider. (It is listed in the KIC with half this orbital period but if it had that orbital period, we would see a transit in Q16 but we do not.) We do not include KOI-872.01 (Kepler-46 b) because \citet{k14742012N} constrain its eccentricity to be below 0.02 based on TTVs. or KOI-1574.01 (Kepler 87-b) because \citet{2013OD} constrain its eccentricity to be below 0.1 based on TTVs. We exclude KOI-3663.01 from our analysis, a.k.a. PH2 \citep{2013W}, because \citet{2013W} measure its eccentricity to be $e = 0.41^{+.08}{-0.29}$; thus it is not super-eccentric. 

\section{Computing the Posterior of the Mean Number of Planets Based on the Observed Number of Planets}
\label{app:poisson}
The probability of observing $\Npl$ from a Poisson distribution with mean $\barNpl$ is:
\begin{equation}
\label{eqn:ppoisson}
{\rm prob}(\Npl | \barNpl) =  \frac{ \barNpl^{\Npl } }{\Npl!} \exp[- \barNpl]
\end{equation}

We wish to determine the posterior distribution for $\barNpl$, give the observed $\Npl$. Applying Bayes' theorem:
\noindent
\begin{equation}
{\rm prob}(\barNpl) | \Npl )  = {\rm prob}(\Npl | \barNpl) {\rm prob}( \barNpl) =  \frac{ \barNpl^{\Npl } }{\Npl!} \exp[- \barNpl] {\rm prob}( \barNpl)
\end{equation}
\noindent where ${\rm prob}( \barNpl)$ is the prior on $\barNpl$. For a uniform prior on ${\rm prob}( \barNpl)$, the median of the posterior, Med($\barNpl$) is the solution to the equation:
\begin{equation}
0.5 = \frac{ \int_{{\rm Med}(\barNpl)}^\infty \frac{ \barNpl^{\Npl } }{\Npl!} \exp[- \barNpl] {\rm prob}( \barNpl) d\barNpl}{ \int_0^\infty \frac{ \barNpl^{\Npl } }{\Npl!} \exp[- \barNpl] {\rm prob}( \barNpl) d\barNpl}
\end{equation}
\noindent
For a uniform prior ${\rm prob}( \barNpl) \propto 1$,
\begin{equation}
\label{eqn:count}
0.5 = \frac{ \int_{{\rm Med}(\barNpl)}^\infty  \barNpl^{\Npl } \exp[- \barNpl] d\barNpl}{ \int_0^\infty \barNpl^{\Npl } \exp[- \barNpl]  d\barNpl} = \frac{\Gamma[\Npl+1,{\rm Med}(\barNpl)]}{\Gamma[\Npl+1]}
\end{equation}
\noindent where $\Gamma$ is the gamma function.
\noindent
For a Jeffrey's prior (appropriate when the scale of the parameter is unknown), which for a Poisson distribution is ${\rm prob}( \barNpl) \propto (\barNpl)^{-1/2}$ (e.g. \citealt{2000B,2013FG,2013TF}),
\begin{equation}
\label{eqn:countj}
0.5 = \frac{ \int_{{\rm Med}(\barNpl)}^\infty  \barNpl^{\Npl -0.5} \exp[- \barNpl ] d\barNpl}{ \int_0^\infty \barNpl^{\Npl  -0.5} \exp[- \barNpl]  d\barNpl} = \frac{\Gamma[\Npl+0.5,{\rm Med}(\barNpl)]}{\Gamma[\Npl+0.5]}
\end{equation}
\noindent
See Section 3.4 of \citet{2013FG} for a detailed discussion of the appropriate prior for Poisson counting statistics. The 68.3\% confidence interval is calculated by equating the ratios in Equation \ref{eqn:count} and \ref{eqn:countj} to 0.1585 and 0.8415 (i.e. .5 $\pm$ 0.683/2). These posteriors have medians slightly larger than the counted numbers because of the skewed shape of a Poisson distribution at small values of the mean $(\barNpl <10)$. It is more probable that we are observing fewer planets than the true mean number than vice versa. As an extreme example, if the mean number of planets per sample is greater than 0, there is some possibility that our sample will happen to contain 0.

\section{Short Period Planet Sample}
\label{app:sampleshort}

Here we list the planets from non-\kep surveys that make up the sample in Table \ref{tab:count}:
\begin{itemize}
\item Moderately eccentric Jupiters ($\Nmodz$) with $2.8 < \pfinal < 5$ days: CoRoT-16-b, HAT-P-2-b, HAT-P-21-b, HAT-P-31-b, HAT-P-34-b, and XO-3-b.
\item Moderately eccentric Jupiters ($\Nmodz$) with $5 < \pfinal < 10$ days: CoRoT-10-b, HAT-P-17-b, WASP-8-b, HD-185269-b,HD-118203-b, HD-162020-b, HD-108147-b
\item Non-\kep Jupiters with $2.8 < P < 5$ days ($\Nppz$): CoRoT-12-b, CoRoT-13-b, CoRoT-17-b, CoRoT-19-b, CoRoT-23-b, CoRoT-5-b, HAT-P-1-b, HAT-P-12-b, HAT-P-13-b, HAT-P-19-b, HAT-P-20-b, HAT-P-21-b, HAT-P-22-b, HAT-P-24-b, HAT-P-25-b, HAT-P-27-b, HAT-P-28-b, HAT-P-30-b, HAT-P-33-b, HAT-P-35-b, HAT-P-38-b, HAT-P-39-b, HAT-P-4-b, HAT-P-8-b, HATS-1-b, KELT-2 A-b, OGLE-TR-211-b, WASP-10-b, WASP-11-b, WASP-15-b, WASP-16-b, WASP-21-b, WASP-22-b, WASP-25-b, WASP-29-b, WASP-31-b, WASP-34-b, WASP-35-b, WASP-37-b, WASP-41-b, WASP-42-b, WASP-45-b, WASP-47-b, WASP-55-b, WASP-6-b, WASP-61-b, WASP-62-b, WASP-63-b, WASP-67-b, WASP-7-b, WTS-1-b, XO-3-b, tau-boo-b, 51 Peg-b, HD-102195-b, HD-149143-b, HD-149026-b, HD-179949-b,  HD-187123-b, HD-209458-b, HD-2638-b, HD-330075-b, HD-63454-b, HD-75289-b, HD-83443-b, HD-88133-b, upsilon And-b, P-201-b, and BD-103166-b. We do not include the following planets because their eccentricities are poorly constrained (and therefore may either be above or below $e = 0.2$: CoRoT-11-b, HAT-P-3-b, HAT-P-9-b, OGLE-TR-10-b, OGLE-TR-111-b, OGLE-TR-182-b, TrES-1-b, TrES-4-b, WASP-13-b, WASP-39-b, WASP-58-b, WASP-60-b, XO-1-b, XO-4-b, and XO-5-b. The following planets have $e$ fixed at 0 in the OD but are constrained in the literature to be below $e<0.2$ and therefore we include them in the sample: HAT-P-1-b, HAT-P-4-b, HAT-P-8-b, HAT-P-12-b, HAT-P-27-b, HAT-P-39-b, OGLE-TR-211-b, KELT-2-Ab, WASP-7,  WASP-11-b, WASP-15-b, WASP-21-b, WASP-25-b, WASP-31-b, WASP-35-b, WASP-37-b, WASP-41-b, WASP-42-b, WASP-47-b, WASP-61-b, WASP-62-b, WASP-63-b, and WASP-67-b. Finally, we constrain the eccentricities of CoRoT-13-b \citep{2010C}, CoRoT-17-b \citep{2011CMD}, and WASP-16-b \citep{2009L} to be $e < 0.2$ by fitting the radial-velocity data, following \citet{metal2010D} except using the MCMC algorithm (instead of the Levenberg-Marquardt algorithm) to obtain posteriors for the eccentricities.
\item \kep Jupiters with $2.8 < P < 5$ days ($\Npp$): KOI-10.01, KOI-17.01, KOI-18.01, KOI-20.01, KOI-97.01, KOI-127.01, KOI-128.01, KOI-135.01, KOI-186.01, KOI-188.01, KOI-195.01, KOI-199.01, KOI-201.01, KOI-204.01, KOI-217.01, KOI-421.01, KOI-611.01, KOI-760.01, KOI-767.01,  KOI-830.01, KOI-838.01,  KOI-908.01, KOI-913.01, and KOI-1074.01. We do not include two known false positives (KOI-208.01 and KOI-895.01, \citealt{2011DS}). We do not include KOI-931.01 because it exhibits a very large (1.7 arcsecond) centroid offset during transit, indicating a likely blend/false positive. For the same reason we do not include KOI-214.01 or KOI-813.01. KOI-554.01 was found to be a brown dwarf by \citet{photo2012SD}.
\item Non-\kep Jupiters with $5 < P < 10$ days ($\Nppz$): CoRoT-16-b, CoRoT-4-b, CoRoT-6-b, HAT-P-18-b, HAT-P-2-b, HAT-P-29-b, HAT-P-31-b, HAT-P-34-b, WASP-38-b, WASP-59-b, WASP-8-b, HD-185269-b, HD-118203-b, HIP 14810-b, HD-68988-b, HD-162020-b, HD-217107-b, HD-109749-b.
\item \kep Jupiters with $5 < P < 10$ days ($\Npp$): KOI-22.01, KOI-98.01, KOI-131.01, KOI-200.01, KOI-206.01, KOI-428.01, KOI-466.01, KOI-680.01, KOI-728.01, KOI-774.01, KOI-889.01, KOI-890.01, KOI-929.01,  KOI-1391.01, KOI-1456.01, and KOI-1465.01. We do not include known false positives KOI-425.01 or KOI-607.01 (\citealt{photo2012SD} and see also \citealt{2012M}). We exclude KOI-3627.02 because it appears to be a false-positive. We do not include KOI-1066.01 because it exhibits a very large (3.1 arcsecond) centroid offset during transit, indicating a likely blend/false positive. For the same reason we don't include KOI-1457.01. We exclude KOI-3721.01, KOI-3767.01, and KOI-3771.01 because they were only observed starting in Q10 and have extreme stellar variability, making it highly unlikely that their super-eccentric prototypes would have been detected. We also exclude KOI-3689.01 because it was only observed during Q10.
\end{itemize}

These numbers differ from those in S12 for several reasons:
\begin{itemize}
\item S12 used the same definition that use we use $\Npp$ for the \kep sample but, for the calibration sample, they used the total number of planets observed along the $\pfinal$ track. However, we wish to treat both samples the same and therefore use the same class of object for both: $\Npp$ for the \kep sample and $\Nppz$ for the calibration sample. 
\item We combine non-\kep planets detected by transit surveys and those detected radial-velocity surveys in order to enhance our sample size and get a better estimate for Equation~(\ref{eqn:predn}), rather than treating the two samples separately. This is justifiable because even though the total number of planets along a given angular momentum track has different selection biases for transit vs. radial-velocity, the \emph{fraction} of moderately eccentric Jupiters along a given track should be consistent in the two samples (as S12 found to be the case). The non-\kep transit surveys are not better suited than the radial-velocity surveys for a comparison to the \kep stars, except for the transit probability. As noted by S12, the transit probability is constant along an angular momentum track. Therefore the ratio of planets along different portions of a given track should not differ between radial-velocity vs. transit samples.
\item We impose cuts on stellar effective temperature and surface gravity to restrict our sample to the type of stars with reliable stellar density estimates.
\item Additional non-\kep planets have been discovered and characterized since S12.
\item Some of the \kep hot Jupiter candidates had their status or parameters changed. Of the forty-six hot Jupiters in the \citet{k14742011BKB} sample used by S12, eighteen are no longer included: one had its radius decreased below 8 Earth radii (KOI-214.01), eleven were designated a false positive (KOI-194.01, KOI-552.01, KOI-609.01, KOI-822.01, KOI-840.01, KOI-876.01, KOI-1003.01, KOI-1152.01, KOI-1177.01, KOI-1382.01, KOI-1543.01), three were declared a false positive in the literature (KOI-208.01, KOI-425.01, KOI-895.01), and one (KOI-931.01) was declared a likely false positive above. KOI-410.01 has a radius of 45 Earth radii and KOI-684 of 30 Earth radii, above our upper limit of 22 Earth radii. 
\end{itemize}

\section{Completeness of \kep sample}
\label{app:compare}

\begin{figure}
\begin{centering}
\includegraphics{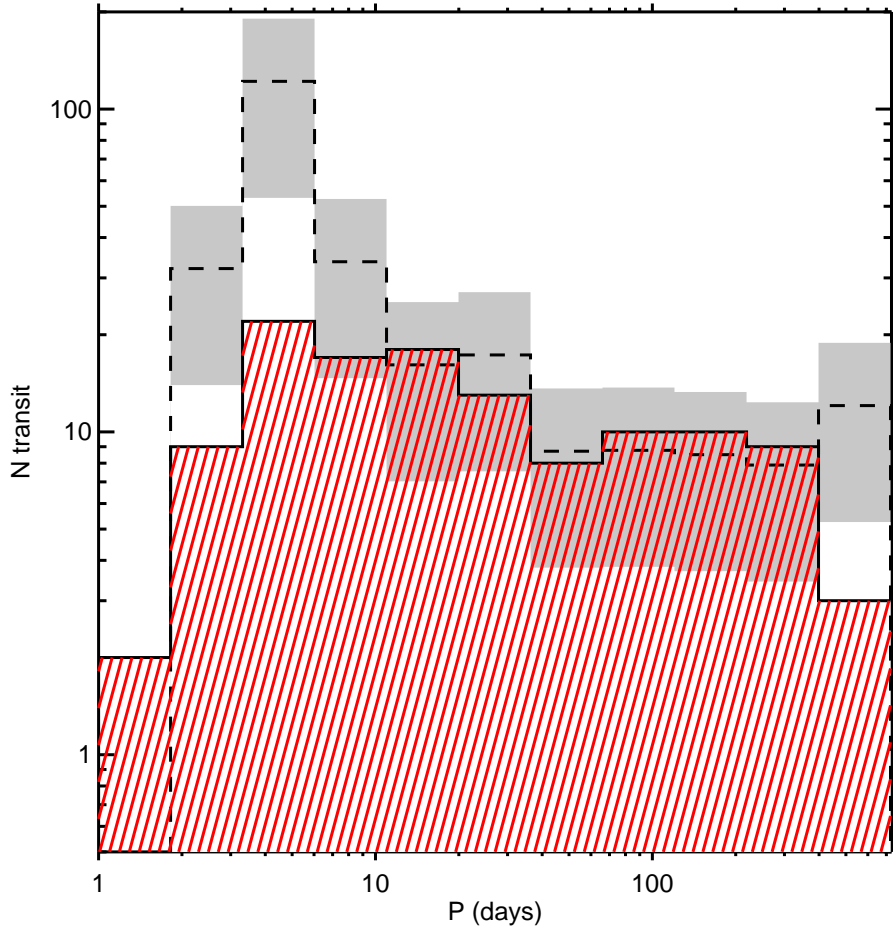}
\caption{Red striped: number of transiting giant planets detected by \kep, using the stellar and planetary cuts described in this work and removing the false positives mentioned in Appendices \ref{app:selectlong} and \ref{app:sampleshort}. Black dashed: expected number based on the RV-discovered (i.e. excluding planets discovered by transit surveys) sample. The gray error bars are from uncertainties in the normalization, not the Poisson uncertainties of each individual bin. \label{fig:compare}}
\end{centering}
\end{figure}

We compare the \kep and radial-velocity samples, following \citet{2013DM} Section 4. We plot (Figure \ref{fig:compare}) the actual observed number of \kep giant planets and compare the number expected from the radial-velocity sample (i.e. those planets plotted in Figure 1), $$N_{\rm RV, trans}=C_{\rm norm}N_{\rm RV}{\rm prob}_{\rm trans},$$ where $N_{\rm RV}$ is the observed number of RV planets per bin and ${\rm prob}_{\rm trans}(P)$ is the transit probability.  We normalize the radial-velocity sample using the absolute occurrence rates of hot Jupiters, which we expect to be complete, and our analytical completeness estimate from Section 2.1. We use the RV hot Jupiter occurrence rate from \citet{k14742012W} ($\fHJRV = 1.2 \pm 0.38 \%$) and the \kep hot Jupiter occurrence rate from \citet{metal2013F} ($\fHJK = 0.43 \pm 0.05 \%$). The normalization constant for the RV sample is:
$$C_{\rm norm}=\frac{\fHJRV}{\fHJK}\frac{\sum_{P = 0}^{10 \rm days}N_{ \rm trans,Kep}(P)/{\rm prob}_{\rm trans}(P)}{\sum_{P = 0}^{10 \rm days}N_{\rm RV} (P) } C_{\rm comp} $$  
\noindent where $ C_{\rm comp}$ is defined in Section 2.1. We note that the comparison here is intended only as a consistency check; a detailed comparison of the \kep and radial-velocity giant planet period distributions would require careful thought about the mapping between mass and radius, the observational biases in both samples (apart from the transit probability), and other considerations.

At the orbital periods considered here $(P>30)$ days, the distributions appear consistent except in the period bin spanning 398-724 days. There are three transiting Jupiters in this bin, whereas from the RV sample we would expect between 5-19 (where the range represents uncertainty in the normalization, not Poisson uncertainties). However, given Poisson uncertainties (for which an observed 3 represents a true number 2-5 at one sigma; see Appendix \ref{app:poisson}), this is consistent with the lower end of the normalization range. Moreover, this period corresponds to the dramatic uptick in planet frequency observed in the radial velocity sample \citep{metal2008C}, and the intrinsic period distribution in the \kep sample may be different than in the radial-velocity sample. We are developing our own pipeline into which we can inject transits to get a better estimate of completeness for a follow-up to this study.

\section{Avoiding Problems Due to Incorrect Stellar Parameters}
\label{app:star}

Characterizing the entire planetary eccentricity distribution from transit light curve parameters can be complicated by systematic errors and uncertainties in the stellar parameters (e.g. \citealt{photo2011M,photo2012KC,photo2012P}). Instead, we simply aim to determine whether or not there are light curves for which $\rhocirc$ is physically unlikely. For example, a planet transiting at periapse with $e =0.95$ would have $\rhocirc = 244 \rho_\star$. Such a high density would be astrophysically implausible based on our knowledge of stellar evolution. 

We consider three potential problems caused by incorrect stellar parameters. First, we could mistakenly identify a planet as being highly eccentric even if it had $\rhocirc \sim \rho_\odot$ if we were to underestimate $\rho_\star$ as being very low. Second, we could miss an eccentric planet if we thought its host star had $\rho_\star \sim \rho_\odot$ but the true stellar density were much smaller. However, we avoid both these problems by restricting our samples to exclude giants.  Regarding the first problem, all of the $\rho_\star$ we derive for \kep hosts in Section 3 are of order 1. Regarding the second problem, by excluding giants from the well-characterized, calibration sample of stars with both transits and RV measurements and from the \kep hot Jupiter sample (Figure~\ref{fig:obs}), we did not make any predictions for super-eccentric planets orbiting giants, and therefore cannot miss any.

The third potential problem results from using the number of hot Jupiters in the \kep sample, $\Npp$, as an input for predicting $\Nsup$ (Equation \ref{eqn:nmod}). If a large fraction of the \kep hot Jupiters orbits stars that are secretly giants but slipped past our stellar parameter cuts, then we might overpredict the expected number of super-eccentric proto-hot Jupiters. As shown by \citet{photo2012MG} and \citealt{photo2013D}, some of the stars classified as M or~K dwarfs might be giants. However, \citet{photo2012MG} find that imposing a cut of $\log g > 4$, as we do, helps avoid this misclassification. With the cut imposed, 97\% of cool stars dimmer than $Kp = 14$ are dwarves \citep{photo2012MG}. Among our sample of \kep hot Jupiters, all the stars with $\teff < 5714$~K have $Kp > 14$, so it is very unlikely our sample harbors many giants masquerading as M or~K dwarves.

Recently, K14 presented several caveats in measuring planet eccentricities using posteriors of $\rhocirc$ and $\rho_\star$. The measurement of $\rhocirc$ from the light curve can be affected by blends, spots, transit timing variations (TTVs), transit duration variations (TDVs), and the planet's mass. K14 shows that a blend factor $B$ reduces the observed $\rhocirc$ by a factor of $B^{-3/4}$.In order to fail to identify a candidate as being supereccentric, the blend would have to be of order $10$. However, this would mean that the observed transit depth has been reduced by a factor of order 3 and that the true transit is not of a planet but of an eclipsing binary star. Therefore, since we are searching for super-eccentric planets not stars, this is not the type of object we are looking for. Conversely, K14 demonstrates that spots cause $\rhocirc$ to be overestimated; therefore this effect would not prevent us from identifying supereccentric planets. We avoid problems caused by TTVs and TDVs by fitting a separate transit time and impact parameter for each transit when justified (e.g. for Kepler-419-b). Finally, K14s show that the planet's mass only affects $\rhocirc$ by $1+M_p/M_\star$, which is small compared to our uncertainties in $\rhocirc$. 

A final caveat by K14 is the limitation of the small-angle approximation used to link the eccentricity $e$ to $\rhocirc$ and $\rho_\star$, e.g. as in DJ12. Condition B (Equation B11) of K14 is that 
\begin{equation}
\label{b11}
\sin x \approx x
\end{equation}
\noindent for which
\begin{equation}
\label{x}
x = \frac{\varrho_c^2}{\sqrt{1-e^2}} \sqrt{ \frac{ (1\pm p)^2 - b^2 }{ (a/R_{\star})^2 \varrho_c^2 - b^2 } }
\end{equation}
\noindent for which (K14, Equation B4):
\begin{equation}
\label{eqn:varrhoc}
\varrho_c = \frac{1-e^2}{1+e\sin\omega}.
\end{equation}

After rewriting Equation (\ref{eqn:varrhoc}) in terms of our variables $g$ and $\afinal$
\begin{equation}
\varrho_c = \frac{\sqrt{\afinal/a}}{g},
\end{equation}
\noindent we can rewrite Equation \ref{x} as:
\begin{eqnarray}
x = \frac{(\frac{\sqrt{\afinal/a}}{g})^2}{\sqrt{\afinal/a}} \sqrt{ \frac{ (1\pm p)^2 - b^2 }{ (a/R_{\star})^2 (\frac{\sqrt{\afinal/a}}{g})^2 - b^2 } }  = \frac{\sqrt{\afinal/a}}{g^2} \sqrt{ \frac{ (1\pm p)^2 - b^2 }{ (\frac{a \afinal}{R_\star^2 g^2}) - b^2 } }
\end{eqnarray}

In order for $\sin x \approx x$, we need $x << \sqrt{6}$ (Equation B13 of K14). For $\sqrt{\afinal/a} < 1$ and $g>1$, we are only in danger of violating this condition when the denominator is small, i.e. $ (\frac{\sqrt{a \afinal}}{R_\star g}) \sim b $. In our search for super-eccentric Jupiters, we consider $\afinal > 8 R_\star$ and $a > 50 R_\star$. Therefore, in order for the LHS to be comparable to an impact parameter (which ranges between 0 and 1), we would need $g > 19 $. We do not have any planets like this in our sample; the biggest $g$ in our sample is of order 2.5.

\section{Exceptional Candidate Treatments}
\label{app:except}

As part of our spectroscopic survey, we measured spectroscopic parameters using {\tt SpecMatch} for KOI-211 and Kepler-419, which we use instead of the KIC parameters. \citet{2013E} conducted a spectroscopic survey of faint KOI. When the \citet{2013HS} are derived from photometry and the \citet{2013E} are derived from spectroscopy, we use the \citet{2013E} parameters instead. This applies to KOI-209, KOI-398, KOI-918, KOI-1089, and KOI-1268. They found that KOI-1288 has $\log g<4$; since this places the star outside the range of stellar parameters we consider, we do not include it in our sample. We also checked if any of the \kep hot Jupiters had their stellar parameters revised outside our range, but none did. 

The \citet{k14742007T} evolution models only include stars with $M_\star > 0.7 M_\odot$. For a subset of low-mass host stars --- KOI-1466, KOI-1477, and KOI-1552 --- we use the Dartmouth stellar evolution models \citep{2008D} instead. We use models with the default helium core mass and solar $\alpha$ abundance. We interpolate/resample to maintain a uniform prior in stellar age, mass, and metallicity.

Based on the V-shapes of their light curves, several of the candidates are probably not planets but grazing eclipsing binaries (KOI-1193.01,KOI-1483.01, KOI-1587.01, KOI-5071.01, and KOI-5241.01). Of the two of these that were candidates in the \citet{metal2013BR} catalogue, both are noted as V-shaped in that catalogue. Although we include them here, we only use the fraction of the posterior for which $R_p < 22 R_\oplus$. Moreover, we use this fraction to weight the contribution to the total number of supereccentric Jupiters in Section \ref{subsec:signif}.

\section{Assumptions that Cannot Explain a Lower than Expected Number of Super-eccentric Proto-hot Jupiters}
\label{app:not}

These assumptions cannot explain observing fewer than expected super-eccentric proto-hot Jupiters, either because a violation would result in \emph{more} super-eccentric progenitors (1) or because they are unlikely to be violated (2-5).

\begin{enumerate}

\item \emph{The evolution of the planet's radius due to tidal inflation is negligible, and no planets are disrupted by tides.} However, if the planet's radius were to expand over the course of HEM due to tidal inflation, then the tidal dissipation rate would be even lower during the earlier stages of HEM, causing planets to spend even longer at high eccentricities. Therefore, this effect could only increase the expected number of super-eccentric proto-hot Jupiters. The prevalence of tidal disruption does not affect the S12 prediction, because the prediction is based on the survivors. Depending on the timescale of tidal disruption, we may observe additional doomed proto-hot Jupiters that will not survive their HEM.

\item \emph{Angular momentum is not exchanged between the planet and star.} If planets were to typically transfer angular momentum to stars, we would expect more super-eccentric hot Jupiters than predicted and vice versa. However, \citet{photo2012PJ} argue that stellar tidal dissipation is likely unimportant, because if it were, most hot Jupiters would be subsumed by their stars on short timescales. We note that although a star can add or remove angular momentum from the planet's orbit as the star rapidly expands on the giant branch (e.g. \citealt{photo2012SD}), we strictly restrict our samples to main-sequence stars so we can ignore this effect. We note that the Sun's spin angular momentum ranges from about 10\% ($\pfinal = $ 3 days) to 6\% ($\pfinal =$ 10 days) the orbital angular momentum of a proto-hot Jupiter.

\item \emph{The planet's orbital angular momentum and spin angular momentum are not exchanged}. We neglect this effect because we assume that the planet maintains a pseudo-synchronous spin throughout its evolution. If the ratio of the planet's orbital angular momentum to its spin angular momentum is large, the planet's spin quickly (compared to the circularization timescale) evolves to this pseudo-synchronous state, in which the planet's spin rate is similar to the orbital frequency at periapse. We expect the ratio of orbital to spin angular momentum is indeed typically large, because the planet's distance from the star is very large compared to the planetary radius.

\item \emph{Moderately-eccentric calibration Jupiters ($\Nmodz$) truly have $e > 0.2$; they are not low-eccentricity planets that appear eccentric due to eccentricity bias.} Eccentricity bias occurs when noise masquerades as eccentricity. Because the eccentricity cannot be negative, it is biased toward higher values. If one decomposes the RV signal caused by an eccentric planet into sinusoidal harmonics of the planet's orbital frequency, one finds that the signal due to eccentricity is primarily embedded in the second harmonic and has an amplitude of $eK$, where $K$ is the RV amplitude (e.g. \citealt{alias2010A}). Eccentricity bias is primarily a concern when $eK$ is near the noise level, i.e. for low-mass and/or long-period planets with small $K$. In contrast, Jupiter-mass planets on short-period orbits have large $K$. For an RV precision of a few m/s and a typical hot Jupiter $K\sim100$ m/s, a signal of amplitude $e\times K = 0.2~\times$~100~m/s~=~20~m/s is well above the noise level. Moreover, an even tighter constraint on the planet's eccentricity is possible through a joint fit to the RVs and transit light curve, as performed for a number of planets in the calibration sample. Therefore we expect that the calibration sample moderately-eccentric Jupiters (which have orbital periods ranging from 3  - 15~days) truly do have $e > 0.2$.

\item \emph{Only a small fraction of \kep hot Jupiters are false positives}. The expected number of proto-hot Jupiters is proportional to the true rate of \kep hot Jupiters $\barNpp$ (Equation \ref{eqn:nmod}). For example, if half the \kep hot Jupiters were false positives, the predicted number of proto-hot Jupiters should be cut in half. \citet{photo2011MJ} and  \citet{2012DC} find low false-positive rates for \kep candidates ($<10\%$). \citet{photo2012SD}, \citet{2012CF}, and \citet{metal2013F} find higher false-positive rates. However, \citet{photo2012SD} focused on a population with high a priori false-positive probabilities because of their V-shaped light curves. Moreover, the false positive discoveries by \citet{2012CF} were for planets with $P < 3$~days, which \citet{2012CF} suggested can be expected from the period distribution of binaries, and we do not include planets with $P<3$~days in our sample here. We have not include the known false-positives in our computations in Section \ref{sec:expect}. Finally, the false positive rate derived by \citet{metal2013F} is somewhat larger (18\%). However, this rate was based on the \citet{metal2013BR} sample and, since then, roughly 25\% of hot Jupiters have been removed from the \kep sample and marked as false positives \citep{metal2013BB}, so we expect the false positive rate of the \citet{metal2013BB} sample that we use is significantly lower.
\end{enumerate}

Finally, S12 assumed that tidal evolution occurs according to the constant tidal time lag approximation \citep{k14741981H,k14741998E,2012SK,2012SKD}. This assumption controls the ratio $r(\emax)$ (Equation \ref{eqn:rC}) of high-eccentricity proto-hot Jupiters to moderate-eccentricity hot Jupiters along a given angular momentum track. The constant tidal time lag approximation is conventional but may be violated: if the dissipation rate were larger for highly-eccentric Jupiters along a given angular momentum track than for moderately-eccentric Jupiters than predicted by the constant time lag model, we would expect fewer super-eccentric proto-hot Jupiters than predicted or vice versa. \citet{2012H} makes the argument that tidal dissipation is stronger at longer orbital periods because the forcing period is longer. However, in the highly eccentric regime we consider here, the forcing period is $\frac{1}{2} \pfinal$ (e.g. \citealt{photo2003W}, Section 3.3). Therefore the forcing period is constant and we expect that if equilibrium tides dominate, the ratio of high-eccentricity to moderate-eccentricity Jupiters used here is not too low.

Dynamical tides, in which dissipation occurs through surface gravity waves (e.g. \citealt{1975Z}), may be important in a proto-hot Jupiter's tidal evolution. \citet{metal2012BN} argue that dynamical tides act at high eccentricities and equilibrium tides at low eccentricities; they added an empirical correction factor to the constant time-lag model so that, at large eccentricities, it matches the numerical results of the dynamical tide model computed by \citet{photo2011I}. The empirical correction factor is proportional to $10^{200 q e^2}$, where $q$ is the periapse distance \citep{metal2012BN}. Along a given angular momentum track, $q = \afinal/(1+e)$, so the tidal dissipation timescale [proportional to $10^{200 a_{\rm final} e^2/(1+e)}$] is longer for larger eccentricities. If this correction factor applies, the contribution of dynamical tides would \emph{increase} the expected number of super-eccentric proto-hot Jupiters. 

The effect of tides on orbital evolution remains uncertain and is a topic of ongoing research. It remains unclear whether the true typical tidal evolution would result in more or in fewer super-eccentric Jupiters. Our results should be revisited as new theories and formulations of tidal evolution are developed.
\clearpage

\bibliography{./paucity} \bibliographystyle{apj}

\end{document}